\documentclass{pasj00}

\begin{document}
\SetRunningHead{Itoh et al.}{Taurus YSOs}
\Received{2000/12/31}
\Accepted{2001/01/01}

\title{Near-Infrared Spectroscopy of Very Low-Luminosity Young Stellar
Objects in the Taurus Molecular Cloud}

\author{Yoichi \textsc{Itoh}}
\affil{Graduate School of Science and Technology, Kobe University,
1-1 Rokkodai, Nada, Kobe, Hyogo 657-8501}
\email{yitoh@kobe-u.ac.jp}

\author{Motohide \textsc{Tamura}}
\affil{National Astronomical Observatory of Japan,
2-21-1 Osawa, Mitaka, Tokyo 181-8588}
\email{tamuramt@cc.nao.ac.jp}
\and
\author{Alan T. \textsc{Tokunaga}}
\affil{Institute for Astronomy, University of Hawaii
2680 Woodlawn Drive, Honolulu, Hawaii 96822, USA}
\email{tokunaga@ifa.hawaii.edu}


\maketitle

\begin{abstract}
We have carried out
near-infrared spectroscopic observations of 23 very low-luminosity young
stellar object (YSO) candidates and 5 their companions in
Heiles Cloud 2, one of the densest
parts of the Taurus molecular cloud.
Twelve objects were confirmed as YSOs by Br$\gamma$ feature.
The effective temperatures of the YSOs and of the companions
are estimated from 
the 2.26 $\micron$ feature, the 2.21 $\micron$ feature, and
the H$_{2}$O band strengths.
Detailed comparisons of our photometric and spectroscopic
observations with
evolutionary tracks on the HR diagram suggest
some objects to be very low-mass YSOs.
\end{abstract}

\section{Introduction}

Recent optical and near-infrared photometric studies have
revealed faint populations of young stellar object (YSO) candidates
in low-mass star forming regions 
(\cite{Comeron}; \cite{Strom95}; \cite{ITG}, hereafter ITG; 
\cite{Barsony}; \cite{Oasa99}),
in intermediate-mass star forming regions
(\cite{Aspin}), and even in high-mass star forming
regions (\cite{Kaifu}; \cite{Lucas00}; \cite{Oasa02}).
Their faintness is suggestive of low-mass; 
they may be very low-mass young stars 
near the stellar/substellar boundary,
young brown dwarfs, or even free floating planets.

From the photometric observations alone, however, it is
impossible to simultaneously determine the mass and age of a YSO.
Near-infrared spectroscopy of YSO candidates is necessary to  
overcome this difficulty (\cite{Greene95}; \cite{Luhman97};
\cite{LuhmanRieke}; \cite{Luhman98}; \cite{Wilking};
\cite{Cushing}; \cite{Lucas01}).
Spectra of some faint YSO candidates exhibit the absorption features
distinctive to late spectral type, implying
young brown dwarfs. 
While such very low-mass objects may be ubiquitous in star-forming
regions, detail of the formation process of such objects, for example
mass function, object density, and disk property, is still unknown.

\citet{Briceno98} have carried out an optical search for very 
low-mass YSOs in the L1495, L1529, L1551, and B209 regions in the 
Taurus molecular cloud. From photometry with 
spectroscopy, they found 9 new YSOs in the clouds, half of them have
very late spectral types, implying very low-mass objects (0.05 \MO -- 0.25 \MO).
\citet{Luhman00} further investigated very low-mass YSOs in the same clouds
by combining optical imaging with near-infrared spectroscopy. They found
that the mass function of the YSOs in these regions has a peak at 0.8 \MO
and is relatively flat between 0.1 \MO and 0.8 \MO range.
In these papers, however, because the targets for the spectroscopy 
were fully or partially selected based on the optical 
color-magnitude diagram,
the sample may not be complete especially for embedded very low-mass
objects.

ITG conducted a near-infrared survey of the central 
\timeform{1D}$ \times$ \timeform{1D} region
of Heiles Cloud 2 in the Taurus molecular cloud, 
one of the best-studied low-mass star forming regions,
with a limiting magnitude of 13.4 mag in the $K$-band.
Fifty YSO candidates
were identified by their intrinsic red color on the ($J-H$, $H-K$) color-color
diagram, following the scheme discussed by \citet{Strom93}.
Successive high-resolution imaging survey discovered
5 companion candidates around the YSO candidates
\citep{ITN}. 
Faintness of some YSOs and their companion may be an 
indicative of the low-mass of the objects.

We describe here near-infrared spectroscopic follow-up of these faint YSO
candidates in the Heiles Cloud 2.
The observations are described in \S 2, and data reduction
procedures in \S 3. 
In \S 4, we derive effective temperatures of the YSOs mainly
from the 2.21 $\micron$ feature, the 2.26 $\micron$ feature, 
and the H$_{2}$O absorption
band. We also calculate photospheric luminosity of the YSOs
from the previous photometry, then plot the YSOs on the HR
diagram. 

\section{Observations}

\begin{table*}
  \caption{The YSO sample in Heiles Cloud 2}\label{HC2pms_spec}
  \begin{center}
\begin{tabular}{lccrrrrcl}
\hline\hline
ITG No. & $\alpha$(1950)& $\delta$(1950)& K & J-H & H-K 
& Av & Obs.\footnotemark[$*$] & Identification\footnotemark[$\dagger$] \\
\hline
2 & \timeform{4h34m57s0} & \timeform{25D53'01''} & 10.05$\pm$0.01 & 0.84$\pm$0.01 & 0.60$\pm$0.01 & 0.14$\pm$0.49 & U \\
 4 & 4~35~15.0 & 26~01~28 & 11.05$\pm$0.02 & 2.07$\pm$0.05 & 1.29$\pm$0.03 & 11.42$\pm$0.55 & U \\
 5 & 4~35~17.1 & 25~46~06 &  8.31$\pm$0.00 & 1.17$\pm$0.01 & 0.72$\pm$0.01 & 3.38$\pm$0.47 & U & Kim 1-19 \\
 6 & 4~35~17.5 & 26~03~19 & 10.37$\pm$0.01 & 0.96$\pm$0.02 & 0.87$\pm$0.02 & 0.14$\pm$0.49 & U & GM Tau \\
 9A& 4~35~56.6 & 25~27~17 & 12.53$\pm$0.03 & 0.88$\pm$0.04 & 0.66$\pm$0.04 & 0.27$\pm$0.42 & S \\
 9B& 4~35~56.6 & 25~27~21 & 14.45$\pm$0.03 & 0.70$\pm$0.04 & 0.70$\pm$0.04 & 0.00$\pm$0.39 & S & separation = \timeform{4".31}  \\
13 & 4~36~15.2 & 25~31~44 & 11.69$\pm$0.02 & 0.84$\pm$0.03 & 0.55$\pm$0.03 & 0.36$\pm$0.51 & U \\
15A& 4~36~40.5 & 25~56~02 & 9.02$\pm$0.00 & 0.86$\pm$0.01 & 0.62$\pm$0.01 & 0.19$\pm$0.47 & U,S \\
15B& 4~36~40.4 & 25~56~06 & 14.50$\pm$0.01 & 1.01$\pm$0.02 & 0.73$\pm$0.02 & 1.48$\pm$0.39 & S & separation = \timeform{2".99} \\
17 & 4~36~43.1 & 25~55~49 & 10.37$\pm$0.01 & 0.96$\pm$0.02 & 0.82$\pm$0.02 & 0.45$\pm$0.49 & U \\
18 & 4~36~46.3 & 25~41~02 & 10.27$\pm$0.01 & 1.96$\pm$0.03 & 1.18$\pm$0.02 & 10.6$\pm$0.50 & U \\
21 & 4~36~57.5 & 25~50~38 & 10.83$\pm$0.01 & 1.49$\pm$0.03 & 0.94$\pm$0.02 & 6.21$\pm$0.51 & U & GKH 5 \\
24 & 4~37~03.6 & 26~01~26 & 13.05$\pm$0.05 & 1.01$\pm$0.09 & 0.70$\pm$0.09 & 1.65$\pm$0.82 & U \\
25A& 4~37~03.8 & 25~59~38 &  8.97$\pm$0.00 & 1.97$\pm$0.01 & 1.37$\pm$0.01 & 9.82$\pm$0.49 & U,S & IRAS 04370+2559\\
25B& 4~37~03.6 & 25~59~36 & 13.38$\pm$0.01 & 2.14$\pm$0.01 & 1.28$\pm$0.01 &12.29$\pm$0.38 & S & separation = \timeform{4".29} \\
27 & 4~37~32.7 & 25~52~34 &  8.46$\pm$0.00 & 1.04$\pm$0.00 & 0.64$\pm$0.00 & 2.28$\pm$0.47 & U \\
28 & 4~37~46.2 & 25~13~16 & 11.87$\pm$0.02 & 0.81$\pm$0.04 & 0.53$\pm$0.04 & 0.09$\pm$0.54 & U \\
29 & 4~37~52.5 & 25~16~19 & 10.74$\pm$0.01 & 0.86$\pm$0.01 & 0.54$\pm$0.01 & 0.63$\pm$0.49 & U \\
33A& 4~38~04.2 & 25~50~23 & 11.50$\pm$0.02 & 1.41$\pm$0.04 & 1.19$\pm$0.03 & 3.94$\pm$0.55 & U,S \\
33B& 4~38~04.1 & 25~50~28 & 16.05$\pm$0.02 & 0.79$\pm$0.04 & 0.54$\pm$0.03 & 0.00$\pm$0.41 & S & separation = \timeform{5".17} \\
36 & 4~38~10.0 & 25~12~27 & 11.01$\pm$0.02 & 0.86$\pm$0.02 & 0.54$\pm$0.02 & 0.63$\pm$0.49 & U \\
39 & 4~38~16.4 & 25~28~24 & 11.80$\pm$0.02 & 1.01$\pm$0.03 & 0.66$\pm$0.03 & 1.88$\pm$0.51 & U \\
40 & 4~38~20.8 & 25~38~10 & 11.47$\pm$0.02 & 2.80$\pm$0.18 & 1.99$\pm$0.05 & 16.68$\pm$0.83 & U & GKH 32 \\
41 & 4~38~34.6 & 25~50~46 &  9.72$\pm$0.01 & 1.66$\pm$0.01 & 1.11$\pm$0.01 & 7.38$\pm$0.49 & U & IRAS 04385+2550\\
43 & 4~38~39.8 & 25~27~30 & 11.24$\pm$0.01 & 0.84$\pm$0.01 & 0.53$\pm$0.01 & 2.97$\pm$0.55 & U \\
45A& 4~38~45.3 & 25~42~37 & 12.77$\pm$0.05 & 0.90$\pm$0.06 & 0.73$\pm$0.07 & 0.21$\pm$0.52 & S \\
45B& 4~38~45.2 & 25~42~37 & 15.76$\pm$0.05 & 1.33$\pm$0.08 & 0.91$\pm$0.09 & 4.42$\pm$0.62 & S & separation = \timeform{2".29} \\
46 & 4~38~52.3 & 25~47~16 & 11.01$\pm$0.01 & 0.83$\pm$0.01 & 0.52$\pm$0.01 & 0.41$\pm$0.49 & U \\
\hline
\multicolumn{9}{@{}l@{}}{\hbox to 0pt{\parbox{180mm}{\footnotesize
\vspace*{2mm}
\footnotemark[$*$] U: UKIRT, S: Subaru
    \par\noindent
\footnotemark[$\dagger$] GKH: \citet{Gomez}, Kim: \citet{Kim}
    }\hss}}
\end{tabular}
\end{center}
\end{table*}

\subsection{UKIRT observations}

Twenty-one YSO candidates were selected for the
spectroscopic observations (table \ref{HC2pms_spec}).
The sample is nearly flux-limited; The $K$-band
magnitudes range from 8 to 12, up to 4 mag fainter than that of
classical T Tauri Stars (CTTSs) in the Taurus molecular cloud.
We missed ITG 8 and ITG 34 probably due to errors in the coordinates.
As template stars,
3 late-type CTTSs, 1 class I source, 
and 7 late-type dwarfs were also observed.

Spectroscopic
observations have been carried out during 1996 November 24 -- 26 
using the UK Infrared Telescope (UKIRT) at the summit of Mauna Kea,
with the Cooled Grating Spectrometer 4
(CGS4, \cite{Mountain}).
CGS4 has a 256$\times$256 InSb array with a spatial scale of 
\timeform{1".22} pixel$^{-1}$.
The 75 line~~mm$^{-1}$ grating was used with 
the \timeform{1".2} slit, providing
wavelength coverage
of 1.8 $\micron$ -- 2.5 $\micron$ with
resolution $R ( = \lambda/\Delta\lambda$)
of 300 at 2.21 $\micron$.
The integration time used for each source was typically 1 -- 20
seconds depending on source brightness. In most
cases, 96 exposures were taken for each object.
Nodding of the telescope was carried out
approximately 16 arcsec along the slit for sky subtraction.
For the YSO candidates and CTTSs, 
SAO 76542 (A2V) was observed for correction of
the effects of telluric absorption.
For late-type dwarfs, we observed stars with spectral types of
A0V -- A3V at similar airmass.
Exposures of an incandescent lamp on and off were taken at the
start of each night as dome flats.
Exposures of a krypton lamp were taken every three or four hours
for wavelength calibrations.

\subsection{Subaru observations}

The observed objects are 5 binary candidates 
listed in \citet{ITN} (table \ref{HC2pms_spec}).
The $K$-band magnitudes of the companions range from 13 to 16.
Six late-type dwarfs were also observed as templates, among which 
3 latest dwarfs were observed both with UKIRT and Subaru.

Spectroscopic
observations have been carried out on 2000 December 4 -- 5 with 
the Infrared Camera and Spectrograph
(IRCS, \cite{Tokunaga98}; \cite{Kobayashi}) on the Subaru Telescope
at the summit of Mauna
Kea, Hawaii. IRCS has a 1024$\times$1024 InSb
array with spatial scale of \timeform{0".058} pixel$^{-1}$.
Typical seeing size was \timeform{0".4} with a stable condition 
for both observing dates, so that all binaries were well separated. 
The grating provided a wavelength coverage of 2.0 $\micron$ -- 2.5 $\micron$. 
The resolving power $R$ was around 350 at 2.2 $\micron$.
The slit width we used is \timeform{0".6},
slightly wider than the seeing size, so that the effective spectral 
resolution might change by seeing size. We measured FWHMs of the 
absorption features in each spectrum for each object, and confirmed 
that the FWHMs do not change in 1 pixel resolution ($\sim$ 6 \AA). Therefore, 
the effective spectral resolution was stable during observations 
of each target.
The integration time used for each source was typically 60 -- 300
seconds depending on source brightness.
8 or 12 exposures were taken for each object with
the telescope dithered
approximately \timeform{7".5} along the slit for sky subtraction.
For the binary candidates, 
SAO 76542 was observed for correction of
the effects of telluric absorption.
For late-type dwarfs, we observed stars with spectral types of
A0V -- A3V at similar airmass.
Exposures of an incandescent lamp on and off were taken at the
end of each night as dome flats.
Exposures of an argon lamp were taken for wavelength
calibrations at the end of each night.

\section{Data Reduction and Results}

The Image Reduction and Analysis Facility (IRAF) software was
used for all data reduction. 
Reduction procedure is similar for both UKIRT and Subaru data.
First, a dithered pair of object frames were subtracted by each other,
then divided by flat fields. 
Next, the data frames were geometrically transformed to
correct the curvature of the slit image caused by the
grating.
The solutions of the geometric transformation and 
the wavelength calibration 
were derived from the spectrum of a 
krypton lamp or argon lamp taken closest in
time to the object.
Then, individual spectra were extracted from the transformed images
using the APALL task.
The region where the
intensity of the object was more than 20 \% of the peak intensity
at each wavelength was summed into a one-dimensional spectrum.
Extracted spectra were then normalized and combined to produce the
final spectra. Low signal-to-noise spectra due 
to the tracking error of the telescope were rejected.
The object spectrum was divided by the standard star
spectrum, 
in which Brackett$\gamma$ absorption line at 2.166 $\micron$
was removed by interpolating across the adjacent continuum with the SPLOT task.
Finally the spectrum was multiplied by a blackbody spectrum of the
temperature appropriate to the spectral type of the standard
star \citep{Tokunaga}.

The $K$-band spectra of the 23 YSO candidates, 5 companion candidates, 
3 CTTSs, a protostar, and
10 late-type dwarfs are shown in figures \ref{itg1} and \ref{dwarf}.
Prominent features in the spectra are the HI Br$\gamma$
(2.17 $\micron$), 2.21 $\micron$ feature,
2.26 $\micron$ feature, CO band (2.29 $\micron$ and longer), 
and H$_{2}$O absorption ($<$2.15 $\micron$ and $>$2.3 $\micron$).

\begin{figure}
  \begin{center}
    \FigureFile(85mm,85mm){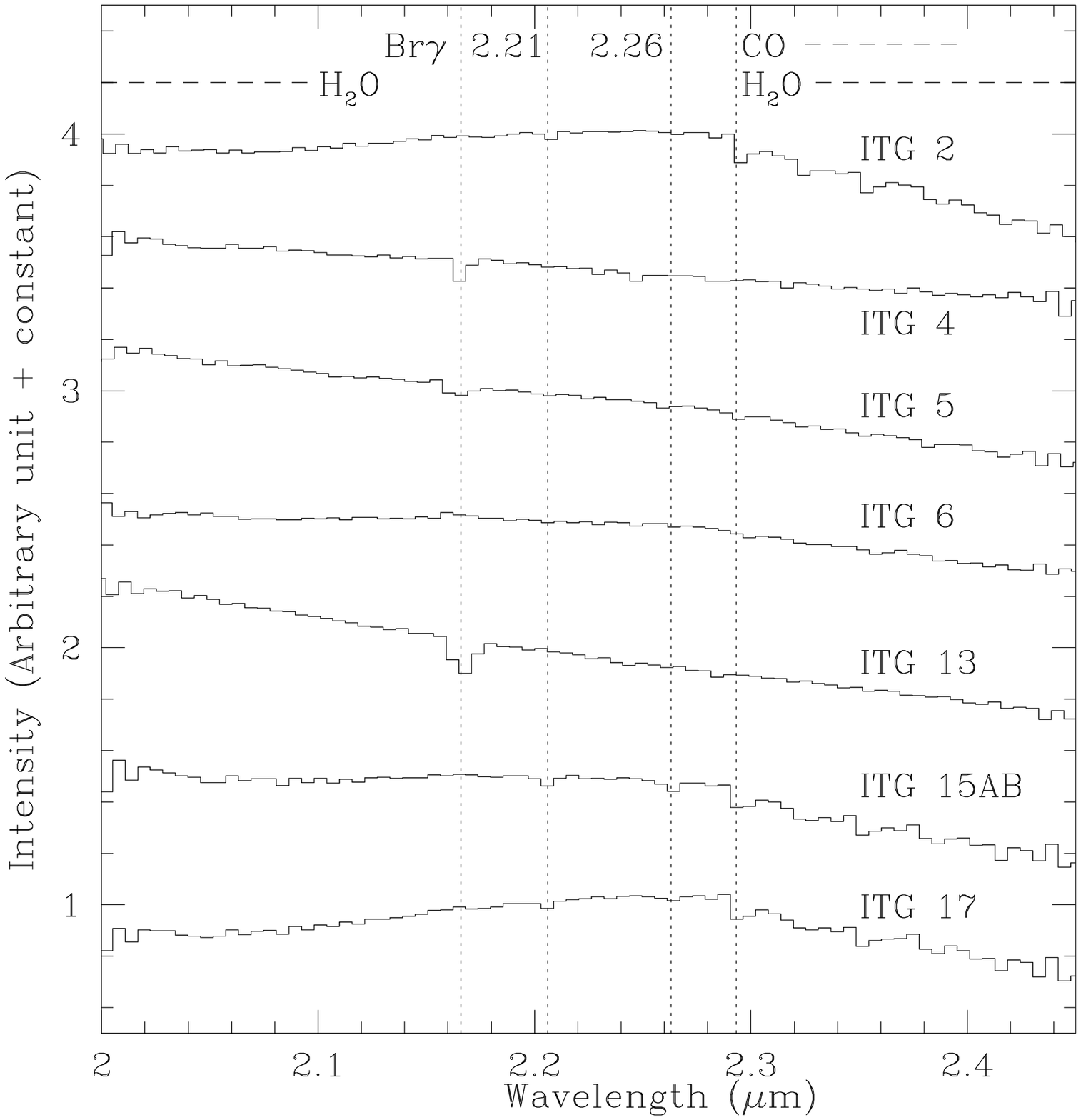}
  \end{center}
  \caption{(a) $K$-band spectra of the YSO candidates taken with UKIRT.
The spectra are normalized by the flux between 2.18 $\micron$ and 
2.20 $\micron$, then offset in steps of 0.5.}
\label{itg1}
\addtocounter{figure}{-1}
\end{figure}

\begin{figure}
  \begin{center}
    \FigureFile(85mm,85mm){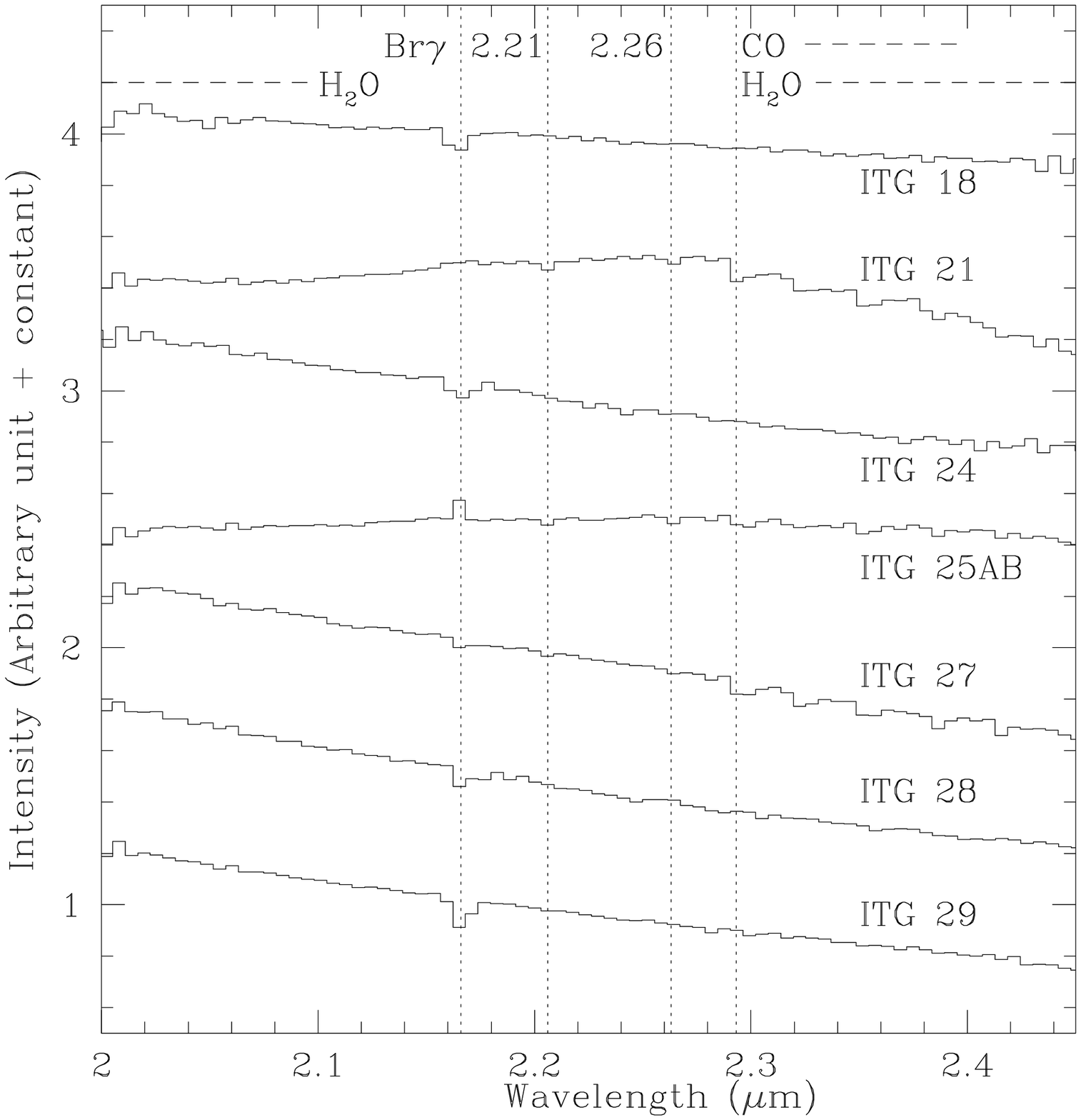}
  \end{center}
  \caption{(b) $K$-band spectra of the YSO candidates taken with UKIRT.
The spectra are offset in steps of 0.5.}
\addtocounter{figure}{-1}
\end{figure}

\begin{figure}
  \begin{center}
    \FigureFile(85mm,85mm){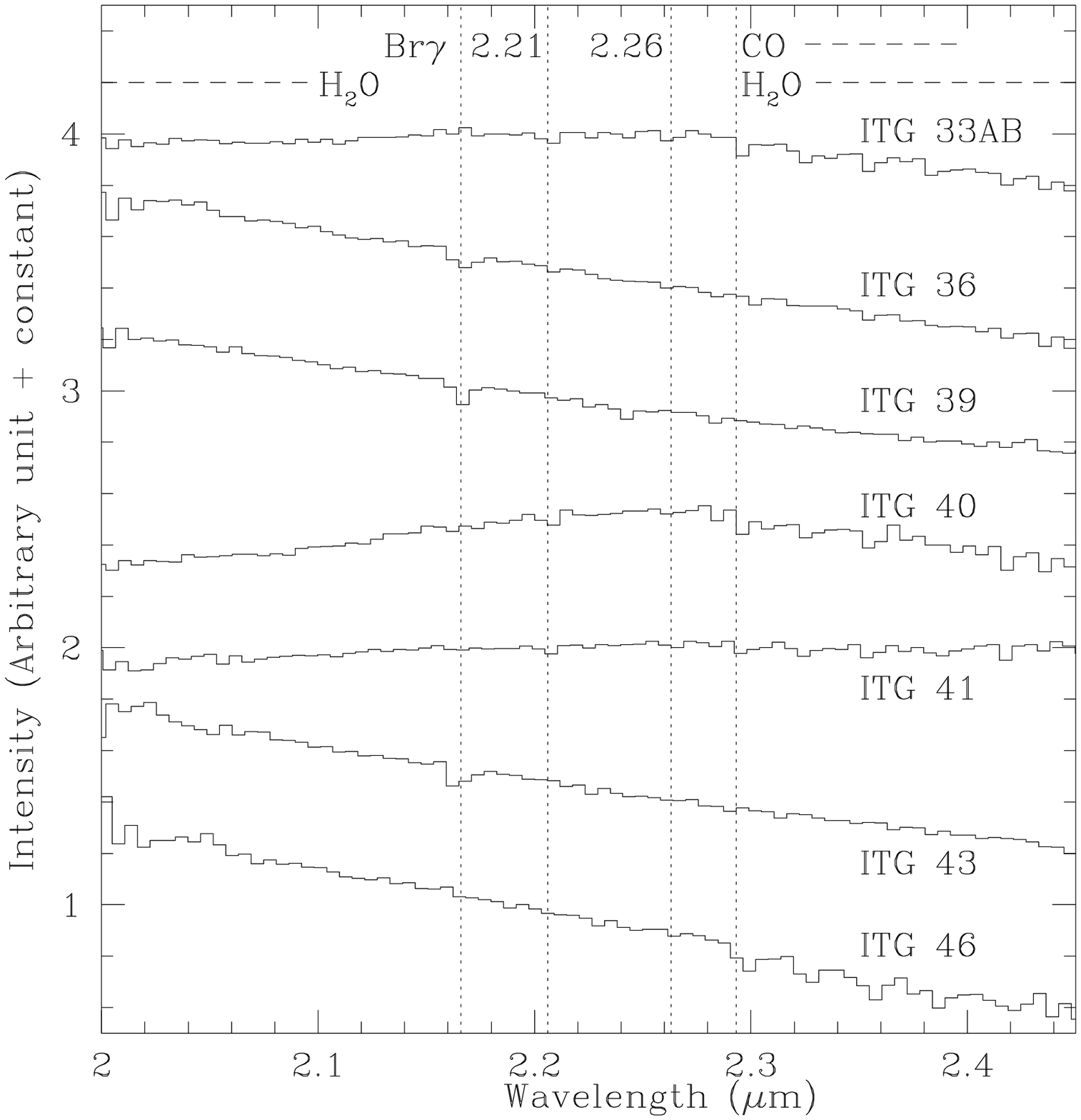}
  \end{center}
  \caption{(c) $K$-band spectra of the YSO candidates taken with UKIRT.
The spectra are offset in steps of 0.5.}
\addtocounter{figure}{-1}
\end{figure}

\begin{figure}
  \begin{center}
    \FigureFile(85mm,85mm){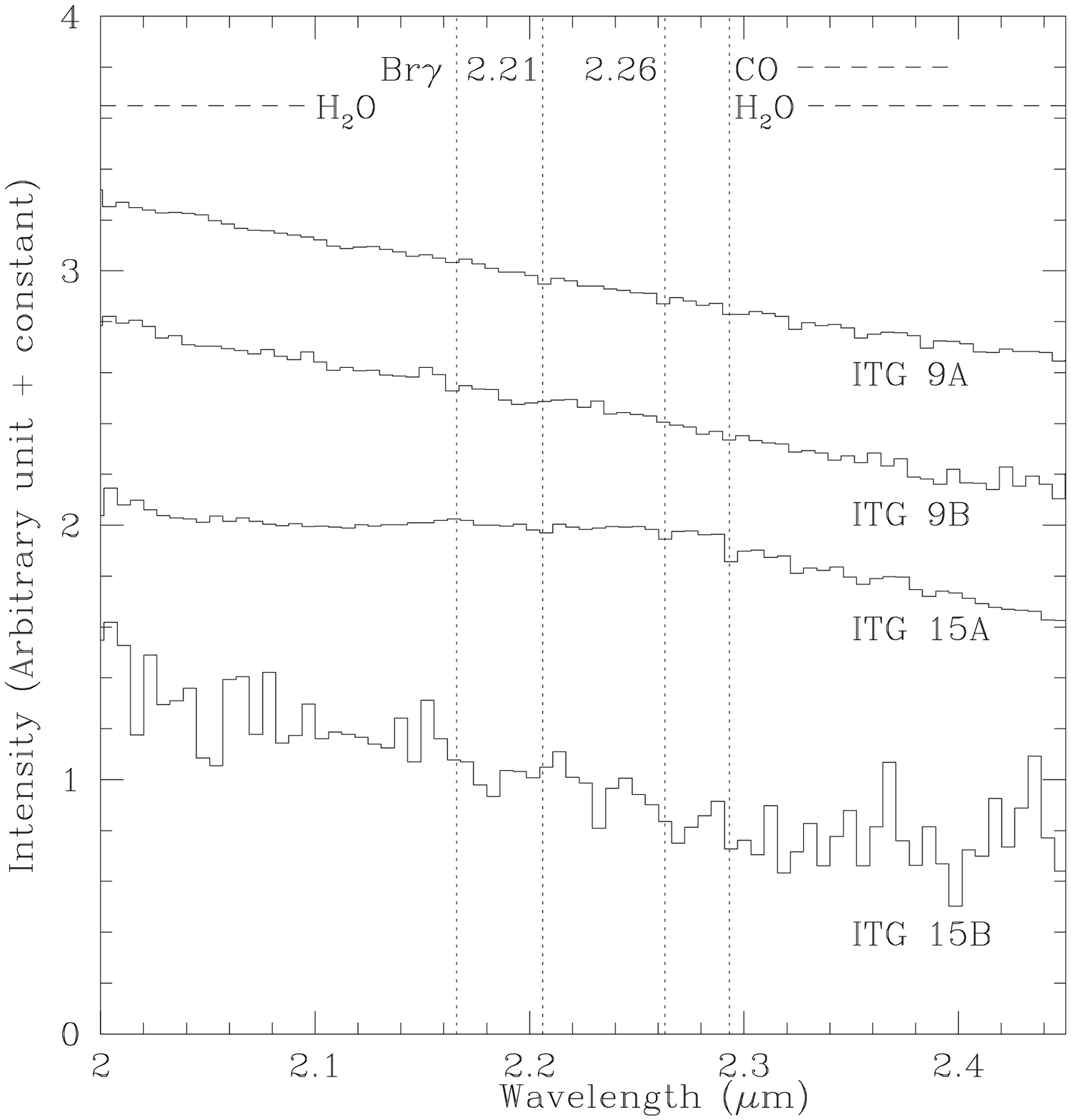}
  \end{center}
  \caption{(d) $K$-band spectra of the YSO binary candidates 
taken with Subaru. Additive constants for the spectra are 0, 1, 1.5, and 2,
respectively.}
\addtocounter{figure}{-1}
\end{figure}

\begin{figure}
  \begin{center}
    \FigureFile(85mm,85mm){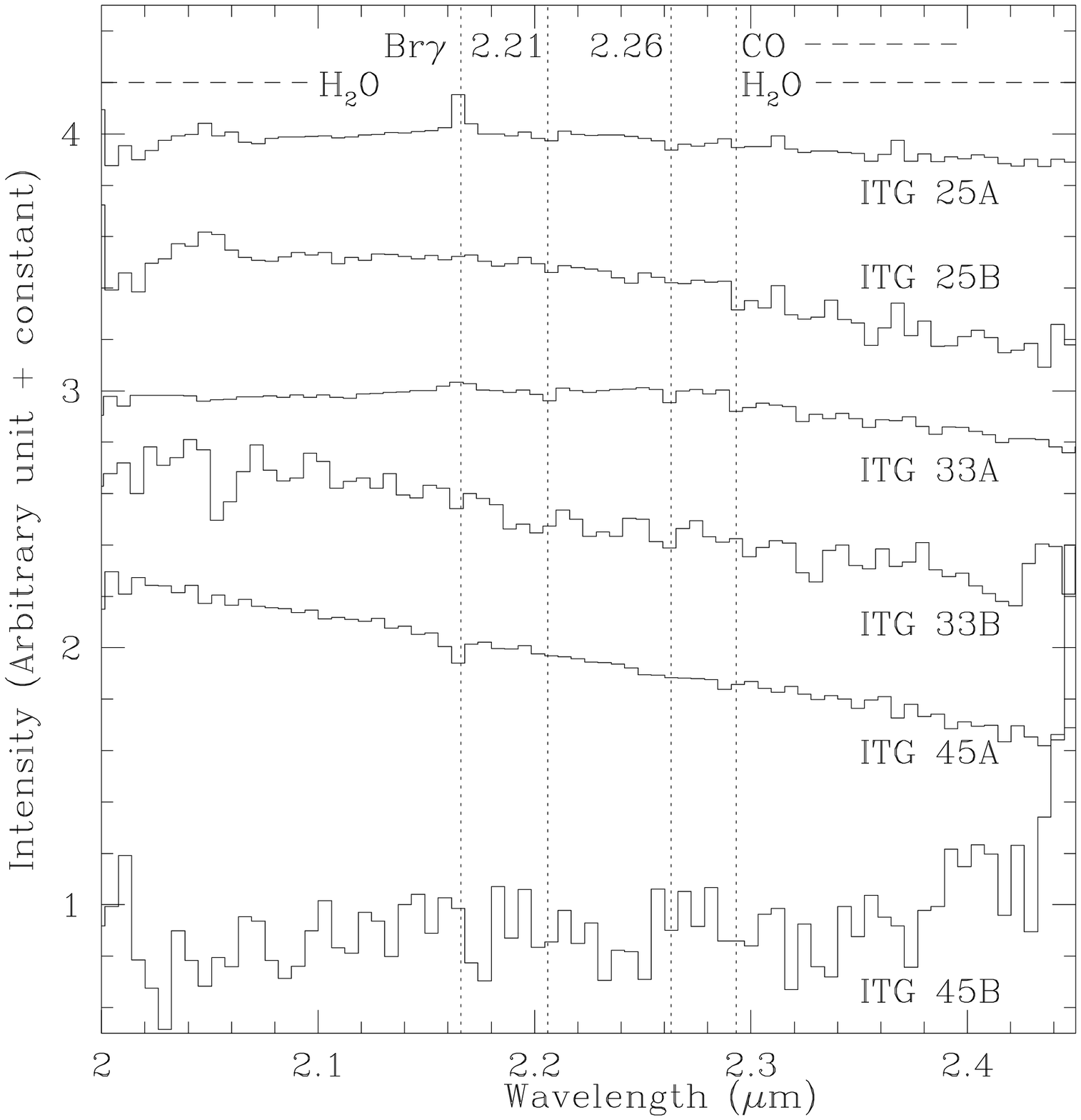}
  \end{center}
  \caption{(e) $K$-band spectra of the YSO binary candidates 
taken with Subaru. Additive constants are 0, 1, 1.5, 2, 2.5, and 3.}
\end{figure}

\begin{figure}
  \begin{center}
    \FigureFile(85mm,85mm){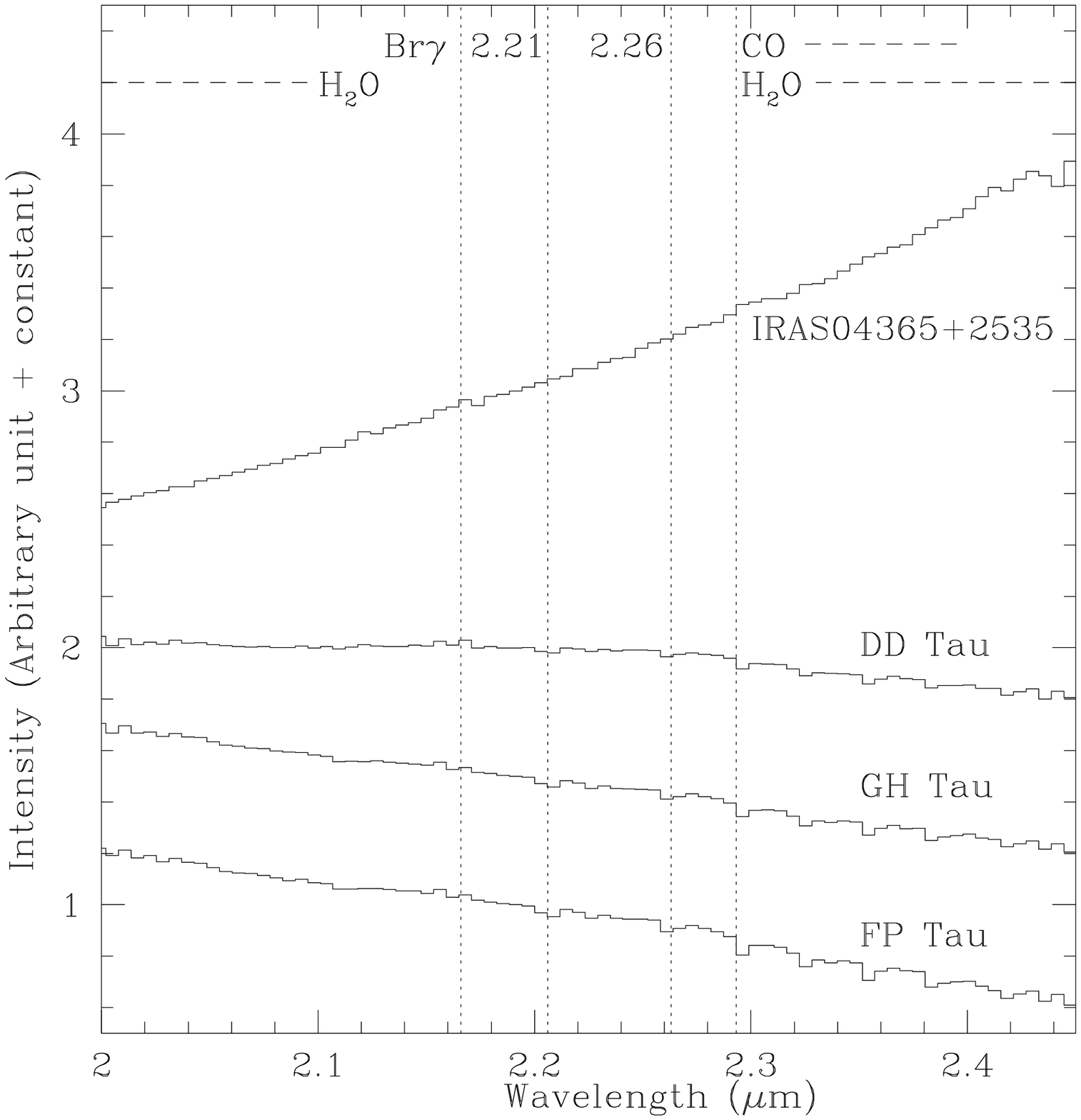}
  \end{center}
  \caption{(a) $K$-band spectra of CTTSs and a protostar taken with UKIRT.
The spectra are normalized by the flux between 2.18 $\micron$ and 2.20 
$\micron$. Additive constants for the spectra are 0, 0.5, 1, and 2, 
respectively.}
\label{tau}
\addtocounter{figure}{-1}
\end{figure}

\begin{figure}
  \begin{center}
    \FigureFile(85mm,85mm){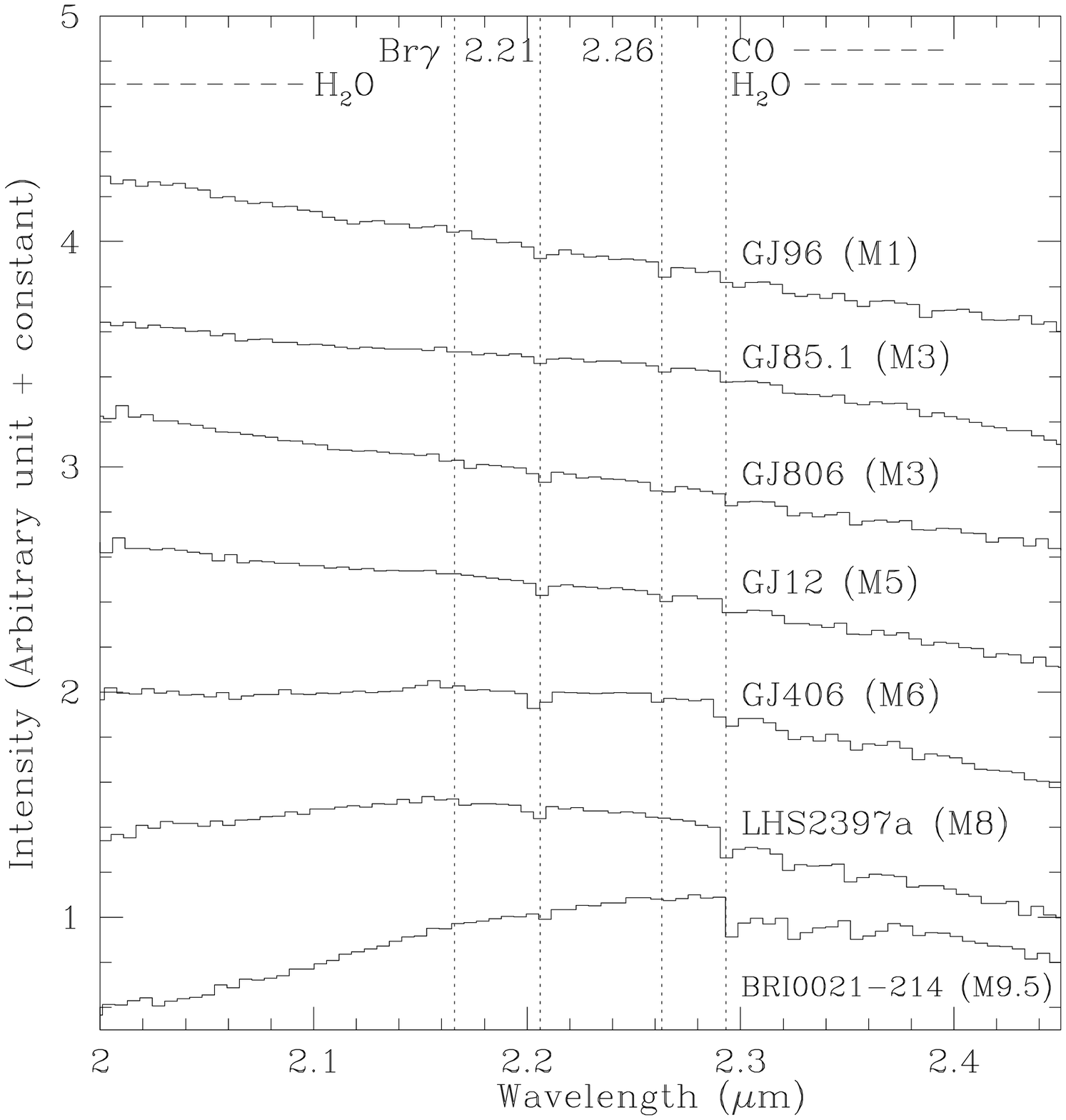}
  \end{center}
  \caption{(b) The $K$-band spectra of late-type dwarfs taken with UKIRT.
The spectra are offset in steps of 0.5.}
\label{dwarf}
\addtocounter{figure}{-1}
\end{figure}

\begin{figure}
  \begin{center}
    \FigureFile(85mm,85mm){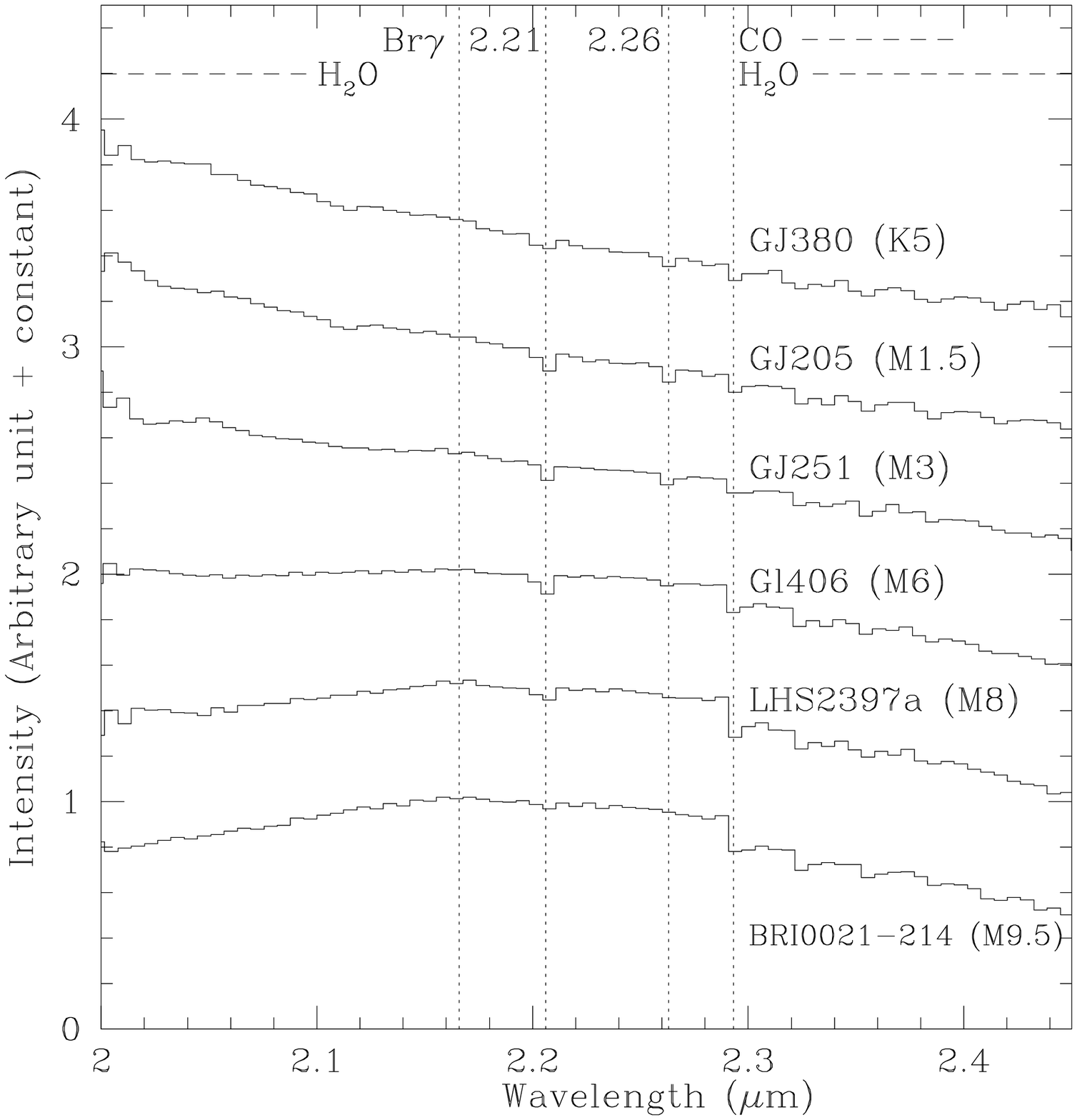}
  \end{center}
  \caption{(c) $K$-band spectra of late-type dwarfs taken with Subaru.
The spectra are offset in steps of 0.5.
}
\end{figure}

Equivalent widths of the Br$\gamma$ line, the 2.21 $\micron$ feature, 
and the 2.26 $\micron$ feature were measured with SPLOT task
by Gaussian fitting. On the other hand, equi\-valent widths of
the CO(2-0) band and the CO(4-2) band
were calculated by simple integration of the absorption intensity.
The uncertainties were estimated from the continuum fit,
in which locating the continuum level was the main factor contributing 
to the equi\-valent width uncertainty. 
We also calculated reddening-independent indices of the H$_{2}$O band
strength $Q$, following \citet{Wilking}. For
Koornneef's extinction law \citep{Koornneef}, the $Q$ index is written as 
\begin{equation}
Q = \left(\frac{F_{1}}{F_{2}}\right)\left(\frac{F_{3}}{F_{2}}\right)^{1.24}
\end{equation}
where $F_{1}$, $F_{2}$, and $F_{3}$ are flux densities between
2.07 $\micron$ -- 2.13 $\micron$, 2.267 $\micron$ -- 2.285 $\micron$, and
2.40 $\micron$ -- 2.45 $\micron$, respectively.

The measured equi\-valent widths of the features and the strengths
of the bands are tabulated
in table \ref{ew_tbl}. 
The signal-to-noise ratio derived from deviations between each
exposure is typically 60 (table \ref{ew_tbl}).
For the latest dwarfs taken both with UKIRT and Subaru,
shapes of the spectra are similar for both observations and most
of the measured equi\-valent widths and band strengths are within
observational uncertainties.

\begin{table*}
  \caption{Equivalent widths of the features and strengths of the band of the
YSO candidates, the CTTSs, and the M dwarfs.}
\label{ew_tbl}
  \begin{center}
    \begin{tabular}{lrrrrrccr}
\hline\hline
    Object & Br$\gamma$ & 2.21 $\micron$ & 2.26 $\micron$ & 
CO(2-0) & CO(4-2) & $Q$ & Obs.\footnotemark[$*$] & S/N\\
& [\AA] & [\AA] & [\AA] & [\AA] & [\AA] \\
\hline
ITG 2  & 0.0$\pm$0.8 & 2.18$\pm$0.39 & 1.00$\pm$0.16
       &  6.20 & 7.80    & 0.51$\pm$0.00 & U & 70\\
ITG 4  & 6.80$\pm$0.25  & 0.0$\pm$0.3 & 0.0$\pm$0.1
        & $\cdots$ & $\cdots$ & 0.98$\pm$0.01 & U & 90\\
ITG 5  & 4.24$\pm$1.50 & 0.80$\pm$0.26 & 1.44$\pm$0.28  
        &  2.71   & 1.97    & 0.83$\pm$0.01 & U & 90\\
ITG 6  & -2.25$\pm$0.62 & 0.98$\pm$0.23 & 0.53$\pm$0.22  
        &  2.30   & 3.79    & 0.81$\pm$0.01 & U & 90\\
ITG 9A & 0.0$\pm$0.5  & 2.42$\pm$0.04 & 2.38$\pm$0.08
        &  3.45   & 4.72    & 0.89$\pm$0.02 & S & 300\\
ITG 9B & 0$\pm$3  & 0$\pm$3 & 0.0$\pm$0.2       
        & 2.71 & 2.35 & 0.88$\pm$0.09 & S & 60\\
ITG 13 & 14.08$\pm$0.70 & 0.0$\pm$0.3 & 0.0$\pm$0.4
        & $\cdots$ & $\cdots$ & 0.94$\pm$0.01 & U & 60\\
ITG 15A\&B & -0.79$\pm$0.79 & 3.29$\pm$1.00 & 4.12$\pm$1.24  
        & 6.73    & 10.39   & 0.64$\pm$0.01 & U & 80\\
ITG 15A& -3.93$\pm$1.08 & 3.29$\pm$0.24 & 3.19$\pm$0.30  
        &  5.88   & 9.79    & 0.61$\pm$0.00 & S & 460\\
ITG 15B& 0.0$\pm$2.5 & 3.53$\pm$2.92 & 14.19$\pm$12.37
        &  25.90   & 26.40    & 1.65$^{+1.41}_{-0.82}$ & S & 9\\
ITG 17 & 0.0$\pm$0.5 & 1.96$\pm$0.37 & 2.48$\pm$0.17  
        & 6.79    & 10.02   & 0.57$\pm$0.01 & U & 30\\
ITG 18 &  8.25$\pm$0.62 & 0.0$\pm$0.3       & 0.0$\pm$0.3        
        & $\cdots$ & $\cdots$ & 0.98$\pm$0.01 & U & 80\\
ITG 21 & -3.8$\pm$3.8  & 2.22$\pm$0.78 & 2.31$\pm$0.44  
        & 5.11    & 8.50    & 0.54$\pm$0.01 & U & 70\\
ITG 24 &  6.60$\pm$1.69  & 0.0$\pm$0.5       & 0.58$\pm$0.12
        & $\cdots$ & $\cdots$ & 1.03$\pm$0.01 & U & 40\\
ITG 25A\&B & -5.34$\pm$0.91 & 1.94$\pm$0.51 & 2.54$\pm$0.50  
        & 3.72    & 5.43    & 0.87$\pm$0.03 & U & 100\\
ITG 25A& -12.63$\pm$0.22 & 3.38$\pm$0.16 & 2.26$\pm$0.08
        & 3.52 & 8.26 & 0.89$\pm$0.02 & S & 290\\
ITG 25B& 0.0$\pm$3.4 & 1.41$\pm$0.60 & 2.53$\pm$1.13
        & 9.61    & 21.31   & 0.73$\pm$0.09 & S & 60\\
ITG 27 & 0.0$\pm$1.2 & 1.36$\pm$0.64 & 1.33$\pm$0.70  
        & 7.87    & 8.50    & 0.86$\pm$0.01 & U & 100\\
ITG 28 &  5.52$\pm$0.72  & 0.75$\pm$0.15  & 0.0$\pm$0.4
        & $\cdots$ & $\cdots$ & 1.00$\pm$0.01 & U & 50\\
ITG 29 & 11.54$\pm$0.80 & 0.47$\pm$0.12          & 0.0$\pm$0.1
        & $\cdots$ & $\cdots$ & 0.96$\pm$0.01 & U & 65\\
ITG 33A\&B & -2.83$\pm$1.05& 3.68$\pm$0.43 & 3.76$\pm$0.37  
        & 6.16    & 8.56    & 0.74$\pm$0.01 & U & 55\\
ITG 33A& -5.85$\pm$0.85 & 3.78$\pm$0.54 & 3.92$\pm$0.47  
        & 6.02    & 7.35   & 0.73$\pm$0.01 & S & 660\\
ITG 33B&  0.0$\pm$2.9 & 7.54$\pm$4.90   & $\cdots$ \        
        & 0$\pm$9 & $\cdots$ & 0.90$\pm$0.04 & S & 20\\
ITG 36 &  5.14$\pm$1.63 & 1.20$\pm$0.43 & 1.26$\pm$0.33  
        & $\cdots$ & $\cdots$ & 0.91$\pm$0.01 & U & 50\\
ITG 39 &  7.40$\pm$0.70 & 0.0$\pm$0.2       &  0.0$\pm$0.2       
        & $\cdots$ & $\cdots$ & 1.00$\pm$0.02 & U & 60\\
ITG 40 & 0.0$\pm$0.9  & 4.15$\pm$1.00 & 1.78$\pm$0.53  
        & 4.95    & 10.76   & 0.67$\pm$0.02 & U & 30\\
ITG 41 & 0.0$\pm$1.0  & 2.31$\pm$0.50 & 2.08$\pm$0.56  
        & 3.43    & 6.53    & 0.93$\pm$0.01 & U & 100\\
ITG 43 &  7.68$\pm$0.98 & 0.0$\pm$0.4       & 0.0$\pm$0.2        
        & $\cdots$ & $\cdots$ & 0.97$\pm$0.01 & U & 60\\
ITG 45A& 8.26$\pm$0.96  & 0.0$\pm$0.88 & 0.0$\pm$1.8
        & $\cdots$    & $\cdots$    & 0.88$\pm$0.07 & S & 80\\
ITG 45B&  0$\pm$25  & 0$\pm$4       &  0$\pm$9        
        & $\cdots$ & $\cdots$ & 1.55$^{+1.77}_{-0.83}$ & S & 6\\
ITG 46 & 0.0$\pm$1.0  & 1.46$\pm$0.50 & 1.34$\pm$0.33  
        & 11.13   & 13.69   & 0.80$\pm$0.01 & U & 70\\
\hline
DD Tau (M1)& -2.53$\pm$0.35 & 1.99$\pm$0.42 & 1.84$\pm$0.25  
        & 3.21 & 4.93 & 0.82$\pm$0.01 & U & 100\\
GH Tau (M2)& -0.58$\pm$0.57 & 2.91$\pm$0.46 & 2.81$\pm$0.48  
        & 5.01 & 6.92 & 0.85$\pm$0.01 & U & 130\\
FP Tau (M5.5)& -1.02$\pm$0.46& $\pm$0.013.35$\pm$0.69& 3.47$\pm$0.38  
        & 6.91 & 8.27 & 0.75$\pm$0.00 & U & 150\\
IRAS 04365+2535 &-1.68$\pm$0.94 & 0.0$\pm$0.2 & 0.0$\pm$0.2 
        & $\cdots$ & $\cdots$  & 1.06$\pm$0.02 & U & 160\\ 
\hline
GJ 380 (K5) & -0.14$\pm$0.25  & 4.20$\pm$0.19 & 4.14$\pm$0.25  
        & 6.28 & 9.56 & 0.93$\pm$0.01 & S & 630\\
GJ 96  (M1) & -1.80$\pm$1.80    & 4.78$\pm$1.18 & 4.24$\pm$0.24  
        & 6.67 & 7.19 & 0.84$\pm$0.01 & U & 110\\
GJ 205 (M1.5)& 0.00$\pm$0.47 & 7.16$\pm$0.51 & 5.80$\pm$0.30  
        & 4.37 & 8.31 & 0.87$\pm$0.01 & S & 780\\
GJ 85.1 (M3)& -1.04$\pm$0.74  & 2.60$\pm$0.60 & 1.97$\pm$0.39  
        & 3.74 & 5.11 & 0.66$\pm$0.01 & U & 310\\
GJ 806 (M3) & -0.40$\pm$0.40    & 4.08$\pm$0.34 & 3.97$\pm$0.36    
        & 4.88 & 7.39 & 0.81$\pm$0.01 & U & 100\\
GJ 251 (M3) & 0.0$\pm$0.1 & 5.67$\pm$0.47 & 4.01$\pm$0.20    
        & 4.90 & 8.08 & 0.77$\pm$0.01 & S & 620\\
GJ 12 (M5)  & 0.0$\pm$0.7    & 4.39$\pm$0.34 & 3.37$\pm$0.27 
        & $\cdots$ & $\cdots$ & 0.71$\pm$0.01 & U & 60\\
GJ 406 (M6) & -1.72$\pm$0.76 & 7.67$\pm$0.75 & 1.65$\pm$0.15  
        & 8.13 & 7.96 & 0.55$\pm$0.01 & U & 60\\
GJ 406 (M6) & 0.0$\pm$0.1 & 7.23$\pm$0.31 & 2.10$\pm$0.95    
        & 6.02 & 7.35 & 0.60$\pm$0.00 & S & 510\\
LHS 2397a (M8)& 0.0$\pm$0.7    & 4.98$\pm$0.31 & 0.39$\pm$0.07  
        & 14.32 & 13.54 & 0.51$\pm$0.01 & U & 90\\
LHS 2397a (M8)& -2.10$\pm$1.19  & 5.04$\pm$0.15 & 0.24$\pm$0.24
        & 11.44 & 12.79 & 0.51$\pm$0.01 & S & 420\\
BRI 0021-214 (M9.5)& 0.0$\pm$0.3    & 2.87$\pm$0.52 & 1.75$\pm$0.82 
        & 11.54 & 10.59 & 0.52$\pm$0.01 & U & 70\\
BRI 0021-214 (M9.5)& 0.0$\pm$0.6 & 2.98$\pm$0.26 & 0.05$\pm$0.02
        & 9.57 & 7.60 & 0.49$\pm$0.01 & S & 200 \\
\hline
\multicolumn{9}{@{}l@{}}{\hbox to 0pt{\parbox{180mm}{\footnotesize
\vspace*{2mm}
    \footnotemark[$*$]
U: UKIRT, S: Subaru
    }\hss}}
    \end{tabular}
  \end{center}
\end{table*}

\section{Discussion}

\subsection{Identification as YSOs using the Br$\gamma$
feature}

We identified 12 sources as YSOs, whose Br$\gamma$ feature is in
emission or in flat (featureless).
We regard a featureless spectrum around the Br$\gamma$ line with 
uncertainty less than 2 \AA ~in equi\-valent width as flat.
The line emission is probably due to accretion of matter from
the circumstellar disk onto the star (e.g. \cite{Najita}).

\begin{figure}
  \begin{center}
    \FigureFile(85mm,85mm){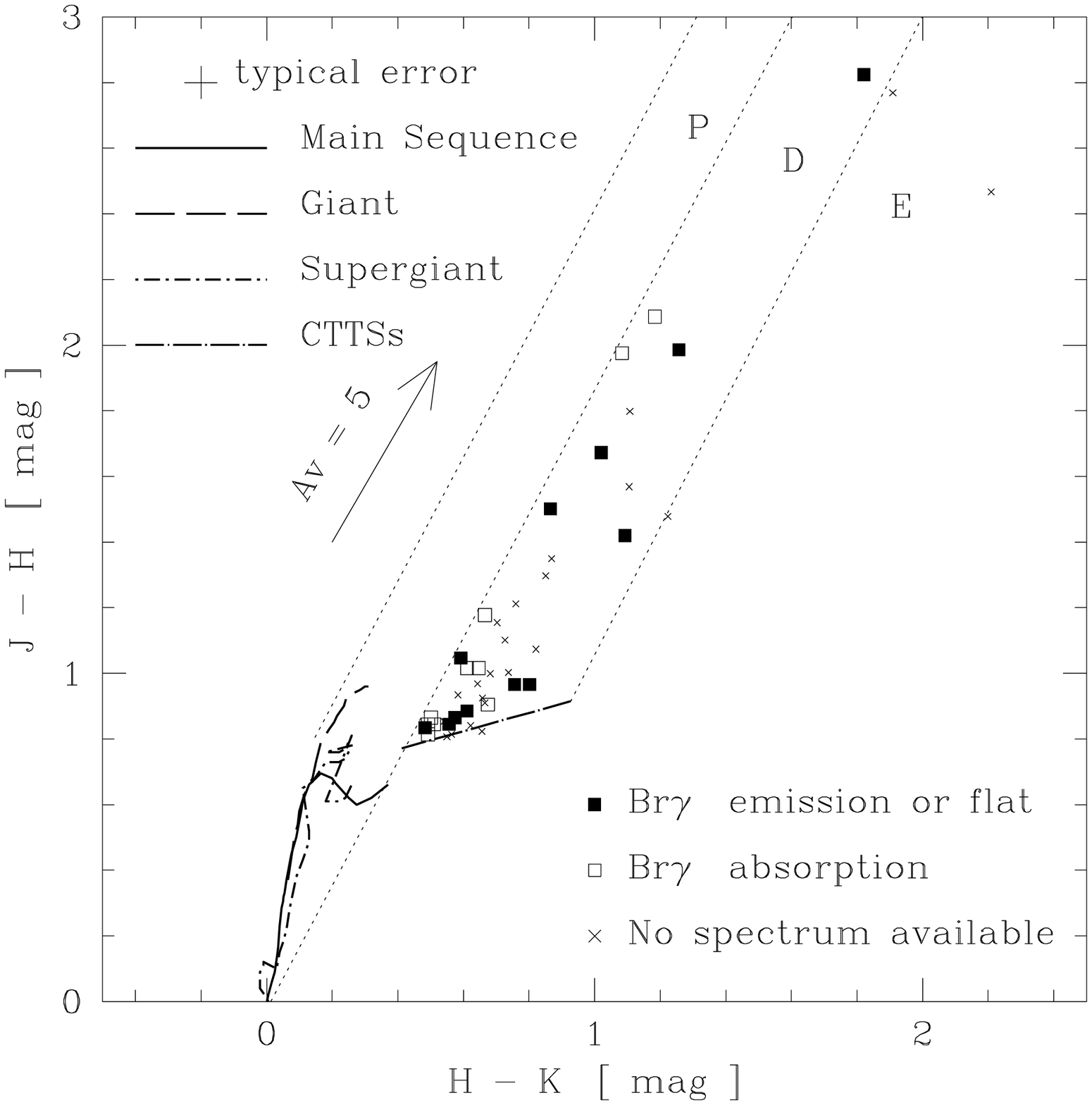}
  \end{center}
  \caption{The color-color diagram for Heiles Cloud 2.
The intrinsic colors of main-sequence stars, giants
\citep{Bessell88}, supergiants \citep{Tokunaga}, and CTTSs
\citep{Meyer97} are plotted with the reddening vector 
(\cite{Koornneef}; E(J-H)/E(H-K)=1.7 in the CIT system).
All colors are transformed to the Johnson/Glass system 
\citep{Bessell88}.
Following the scheme discussed by \citet{Strom93}, the
near-infrared sources are classified into three groups. In the
"P" (photosphere) region on the color-color diagram, located are main 
sequence stars, giants, weak-line T Tauri stars
objects, and the CTTSs whose near-infrared
excess is small. In the "D" (disk) region, a part of CTTSs
are located. Protostars are located in the "E" (envelope) region on
the color-color diagram.
The objects located in the "D" region or the "E" region were
identified as YSO candidates by the previous photometric study (ITG).
The objects without absorption (i.e. with emission or in flat)
in the Br$\gamma$ line are denoted by filled squares, as well as the objects
with absorption by
open squares. The YSO candidates without the spectroscopic observations are
plotted by crosses. 
Field stars identified by ITG are shown by dots.}
\label{ccHC2}
\end{figure}

The other 11 objects with Br$\gamma$ absorption are, on the other hand,
likely early-type field-stars. 
Equivalent widths of the Br$\gamma$ absorption feature are 
4 \AA -- 15 \AA, consistent with the spectral 
types of B, A, and 
early F \citep{Ali}. 
Their locations on the color-color
diagram are close to that of early-type field stars.
Figure \ref{ccHC2} shows the color-color diagram of the
YSO candidates
and field stars identified by ITG toward
Heiles Cloud 2. 
The objects with the Br$\gamma$ absorption
tend to be located near the boundary between the "YSO region" and the 
"field-star region". Therefore, these objects may be early-type
field-stars and have been misidentified as YSO candidates
by photometric classification. 
The star count model of \citet{Jones81} predicts 9 B-
and A-type stars for $K<12$ in a
\timeform{1D} $\times$ \timeform{1D} region toward Heiles Cloud 2. 
This predicted number is consistent with the observational number of the
early type stars.
Note that the number of such early-type stars does not
increase for $K>12$ toward high latitude direction such as the
Taurus molecular cloud.

The uncertainty of the slope of the reddening 
vector may also account for the misidentification.
For the Taurus
molecular cloud, the smallest value of E($J-H$)/E($H-K$) is 1.58
\citep{Elias}, whereas the highest is 2.0 \citep{Gomez}. 
From photometric study alone, identification of YSOs 
depends on the extinction law especially for
the objects near the boundary between
the YSO region and the field star region on the color-color
diagram.

\subsection{Effective Temperature}

We first describe the possible 
$K$-band features with which the effective temperatures of the M
dwarfs and giants are derived. We then apply this method to
the YSOs to derive their effective temperatures.

\begin{figure}
  \begin{center}
    \FigureFile(85mm,85mm){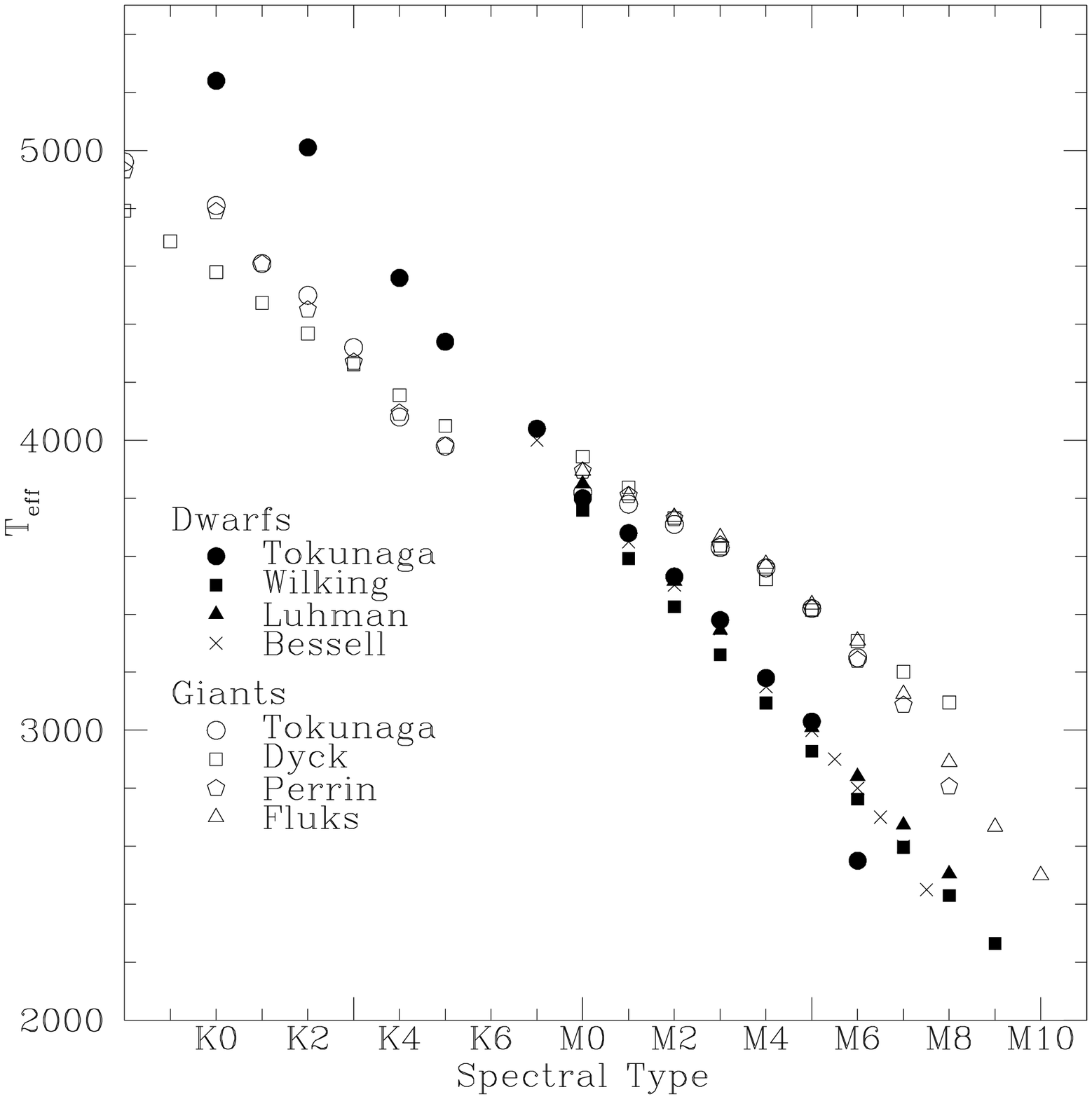}
  \end{center}
  \caption{Relation between effective temperature and spectral type for
dwarfs and giants. Data are taken from \citet{Tokunaga}, \citet{Wilking}, 
\citet{LuhmanRieke}, \citet{Bessell91}, \citet{Dyck}, \citet{Perrin}, 
and \citet{Fluks}.}
\label{SpTeff}
\end{figure}

For cooler objects the conversion from MK spectral type to 
effective temperature is not straightforward because models 
do not exist to give the spectral type vs. effective temperature 
for pre-main sequence stars.  Because the YSOs have luminosity 
class between dwarfs and giants, as we discuss below, we use the 
conversions both for dwarfs and for giants.  In this paper we have 
adopted the conversions given by \citet{Tokunaga} both for dwarfs 
and giants, \citet{Bessell91} for late M dwarfs, and \citet{Fluks} 
for late M giants.  As can be seen in figure \ref{SpTeff}, 
the difference in effective temperature is as much as 800K between 
M dwarfs and M giants even for the same spectral type.

\subsubsection{2.21 $\micron$ feature}

\begin{figure}
  \begin{center}
    \FigureFile(85mm,85mm){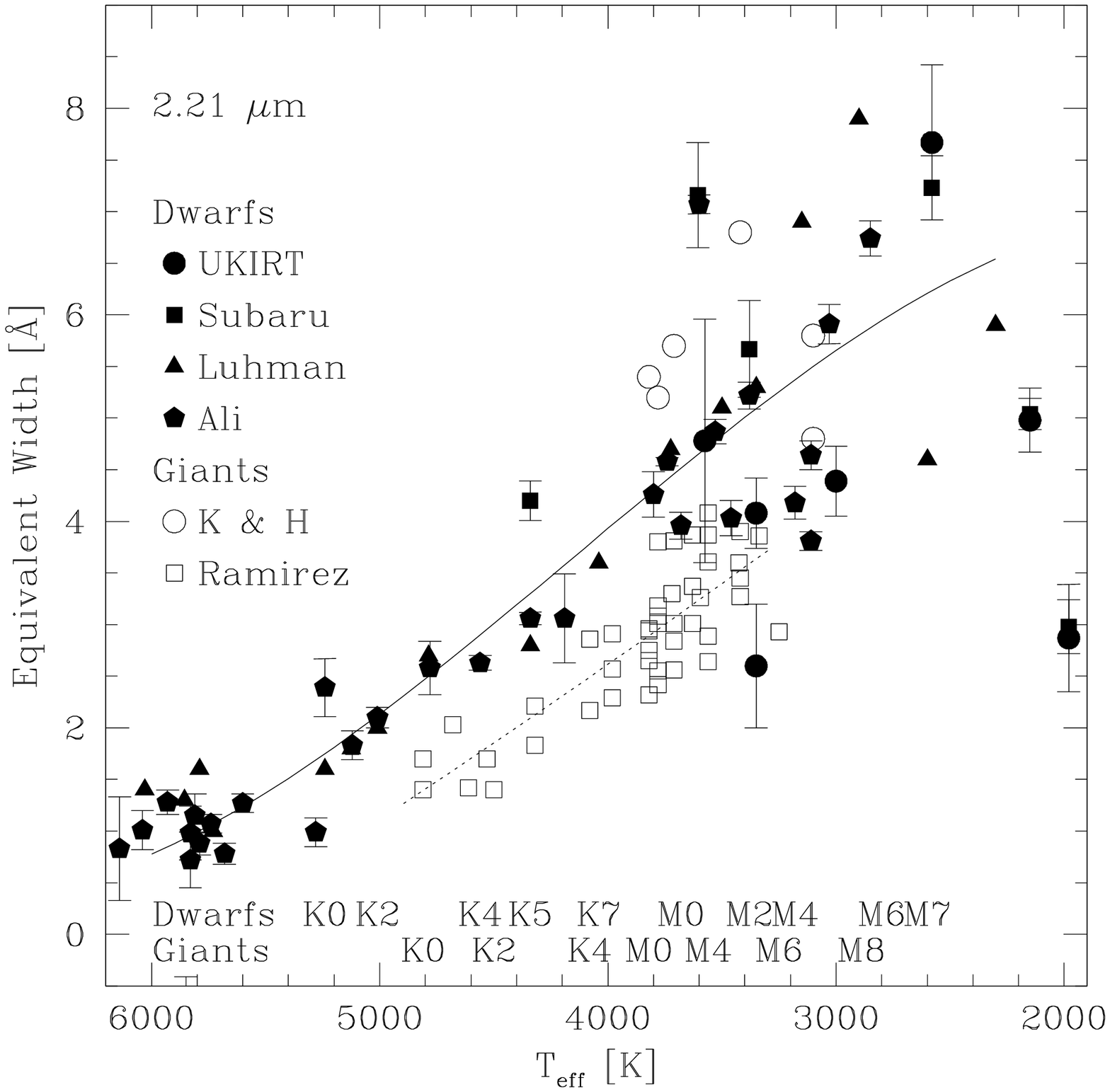}
  \end{center}
  \caption{Equivalent widths of the 2.21 $\micron$ feature
as a function of effective temperature.
Data are taken from \citet{LuhmanRieke}, \citet{Ali}, \citet{Kleinmann}, 
\citet{Ramirez}, and this paper. 
Filled symbols represent the equi\-valent widths for dwarfs and
open symbols for giants.
A solid line shows a fitted line for dwarfs with T$_{\mathrm{eff}} >$ 2300 K, 
and a dotted line for giants listed in \citet{Ramirez}.}
\label{NaTeff}
\end{figure}

The 2.21 $\micron$ feature consists mainly of Na, Sc, Si, and V 
absorption lines
for late-type objects \citep{Ramirez}.
The low excitation energy of these lines
makes it a strong absorption feature in M type stars.
Figure \ref{NaTeff} shows equi\-valent widths of the 2.21 $\micron$
feature of dwarfs and giants as a function of effective temperature. 
The equi\-valent widths increase with decreasing effective temperature.
Moreover, dwarfs have larger equi\-valent widths than giants at the same
effective temperature.
The 2.21 $\micron$ feature has, therefore,
a strong dependence on effective temperature
and weak dependence on gravity
(\cite{Kleinmann}; \cite{Terndrup};
\cite{Ramirez}).

Near-infrared Na lines of CTTSs have symmetric profile, indicating
photospheric origin \citep{Greene97}, while optical Na lines are
red-shifted, indicating outflow origin \citep{Gullbring}.

\subsubsection{2.26 $\micron$ feature}

\begin{figure}
  \begin{center}
    \FigureFile(85mm,85mm){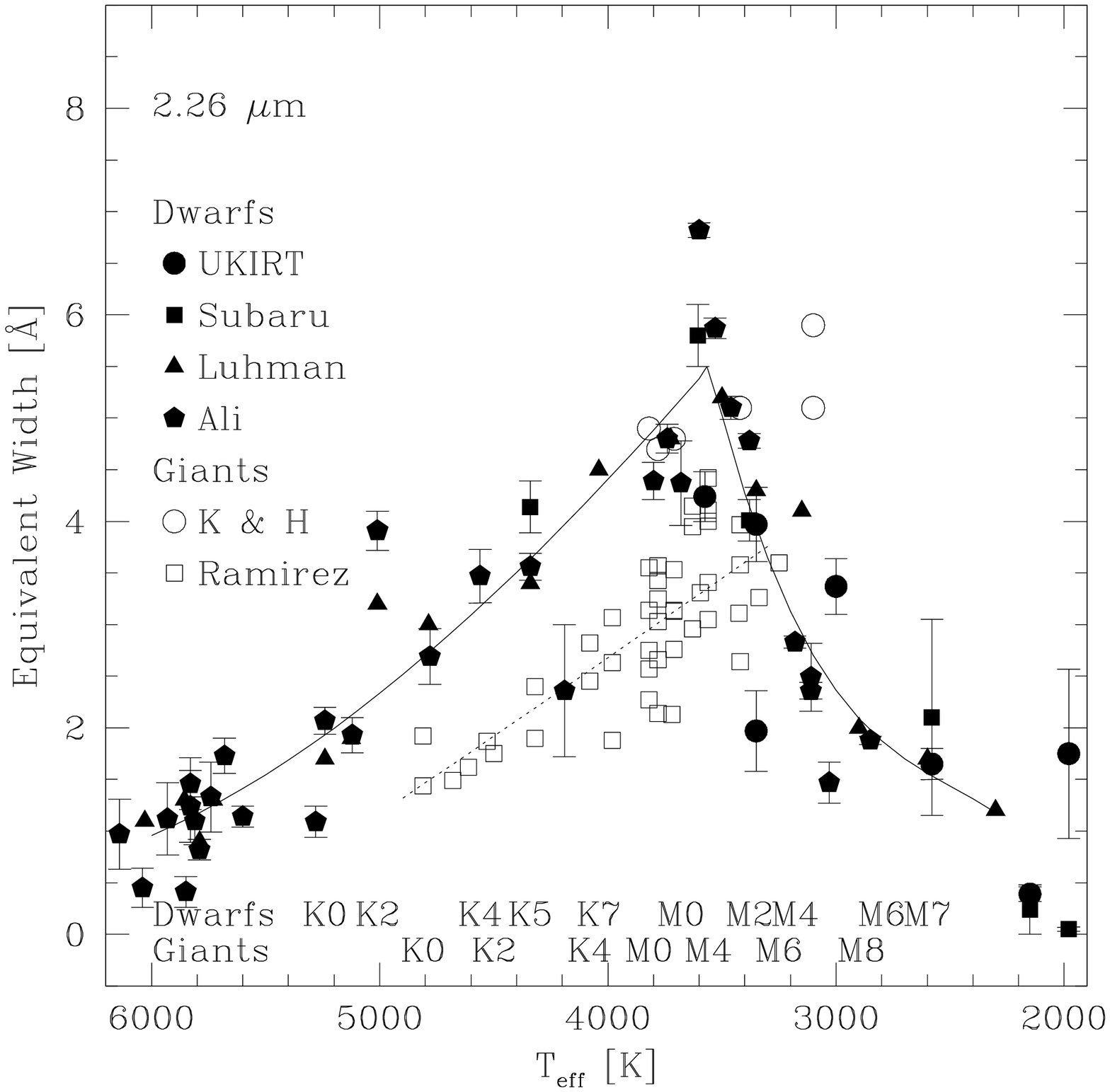}
  \end{center}
  \caption{Equivalent width of the 2.26 $\micron$ feature
as a function of effective temperature.
Filled symbols represent the equi\-valent widths for dwarfs and
open symbols for giants.
Solid lines show fitted lines for dwarfs, and
a dotted line for giants listed in \citet{Ramirez}. 
The intensity has a peak at 3500 K for dwarfs.}
\label{CaTeff}
\end{figure}

The 2.26 $\micron$ feature consists mainly of Ca, Si, and Ti lines
for late-type objects \citep{Ramirez}.
Figure \ref{CaTeff} shows equi\-valent widths of the 2.26 $\micron$
feature of dwarfs and giants as a function of effective temperature. 
The 2.26 $\micron$ feature has a peak in its equi\-valent width around 3500
K, then its width decreases with decreasing temperature, 
due to moderate excitation energy of Ca.
This feature has also strong dependence on effective temperature,
and weak dependence on gravity \citep{Ramirez}. 

\subsubsection{Ratio of Metallic features}

\begin{figure}
  \begin{center}
    \FigureFile(85mm,85mm){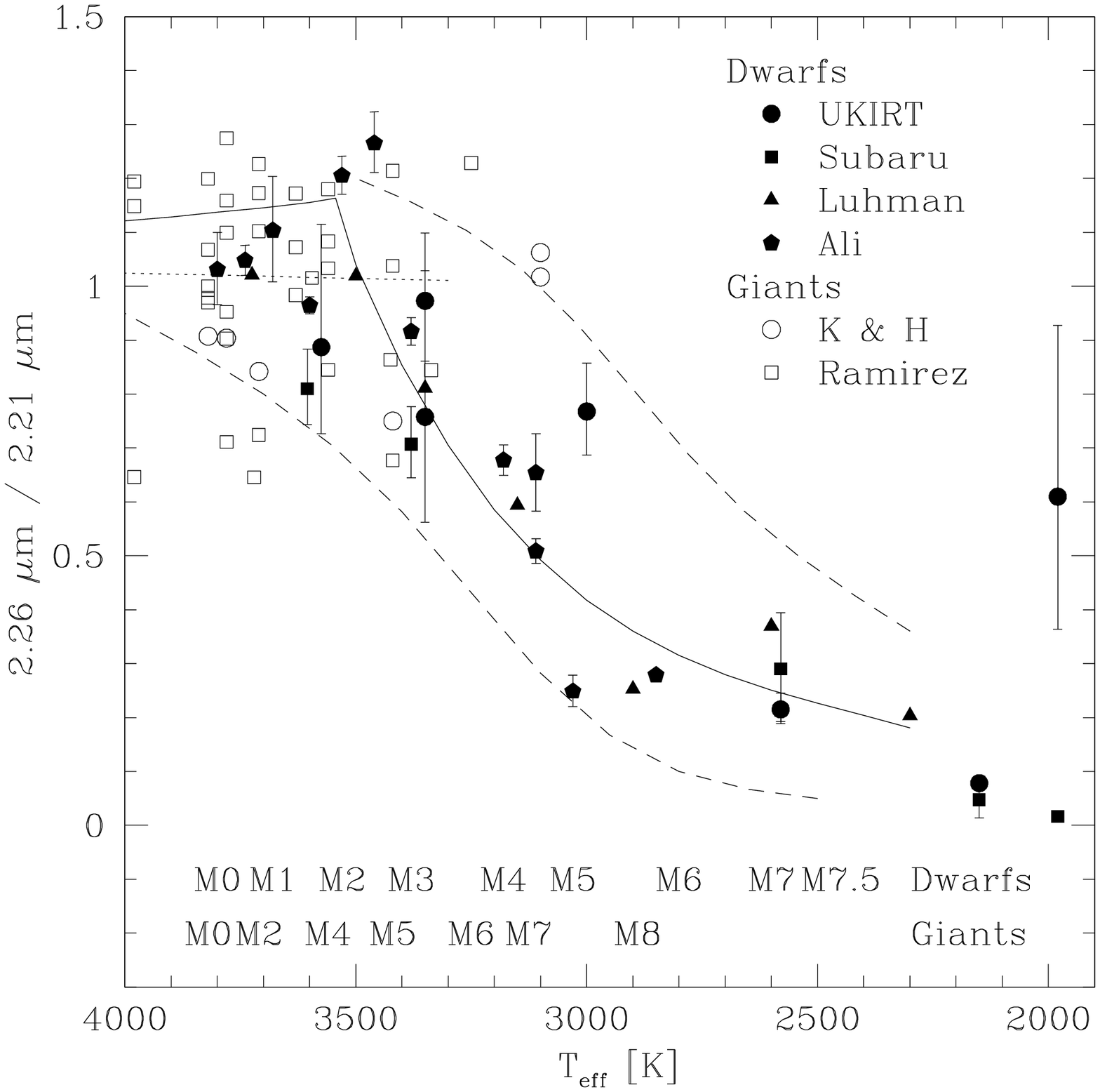}
  \end{center}
  \caption{Ratio of the 2.26 $\micron$ feature to the 2.21 $\micron$ feature
as a function of effective temperature.
Data are taken from our UKIRT observations, Subaru observations,
\citet{LuhmanRieke}, \citet{Ali}, \citet{Kleinmann},
and \citet{Ramirez}.
Solid lines show the ratio for dwarfs, and a dotted line for giants,
both  derived from the fitted lines
in figures \ref{NaTeff} and \ref{CaTeff}.
Dashed lines show the conversions used for estimate of the upper limit
and the lower limit of effective
temperature of the YSOs.}
\label{TCaNa}
\end{figure}

Use of ratio of feature strengths to derive effective temperature
can avoid not only the effect of metallicity
but also the effect of veiling if features
whose wavelengths are close with each other are selected \citep{Meyer96}.
We have checked if the 2.21 $\micron$ feature and the 2.26 $\micron$ feature
can be used as a ratio index of effective temperature.
Figure \ref{TCaNa} shows the 2.26 $\micron$ / 2.21 $\micron$
feature ratio for
dwarfs and giants as a function of effective temperature.
Although the scattering is relatively large and no data is
available for cool giants, the ratio decreases
as the temperature decreases.
Dwarfs and giants have similar values, so that
surface gravity does not appear to significantly affect
this feature ratio.
Therefore, this feature ratio is a sensitive indicator of
effective temperature for T$_{\mathrm{eff}} <$ 3300 K.
For 2.26 $\micron$ / 2.21 $\micron$ $>$ 1, only lower limits of 
effective temperature can be estimated.

\subsubsection{H$_{2}$O}

\begin{figure}
  \begin{center}
    \FigureFile(85mm,85mm){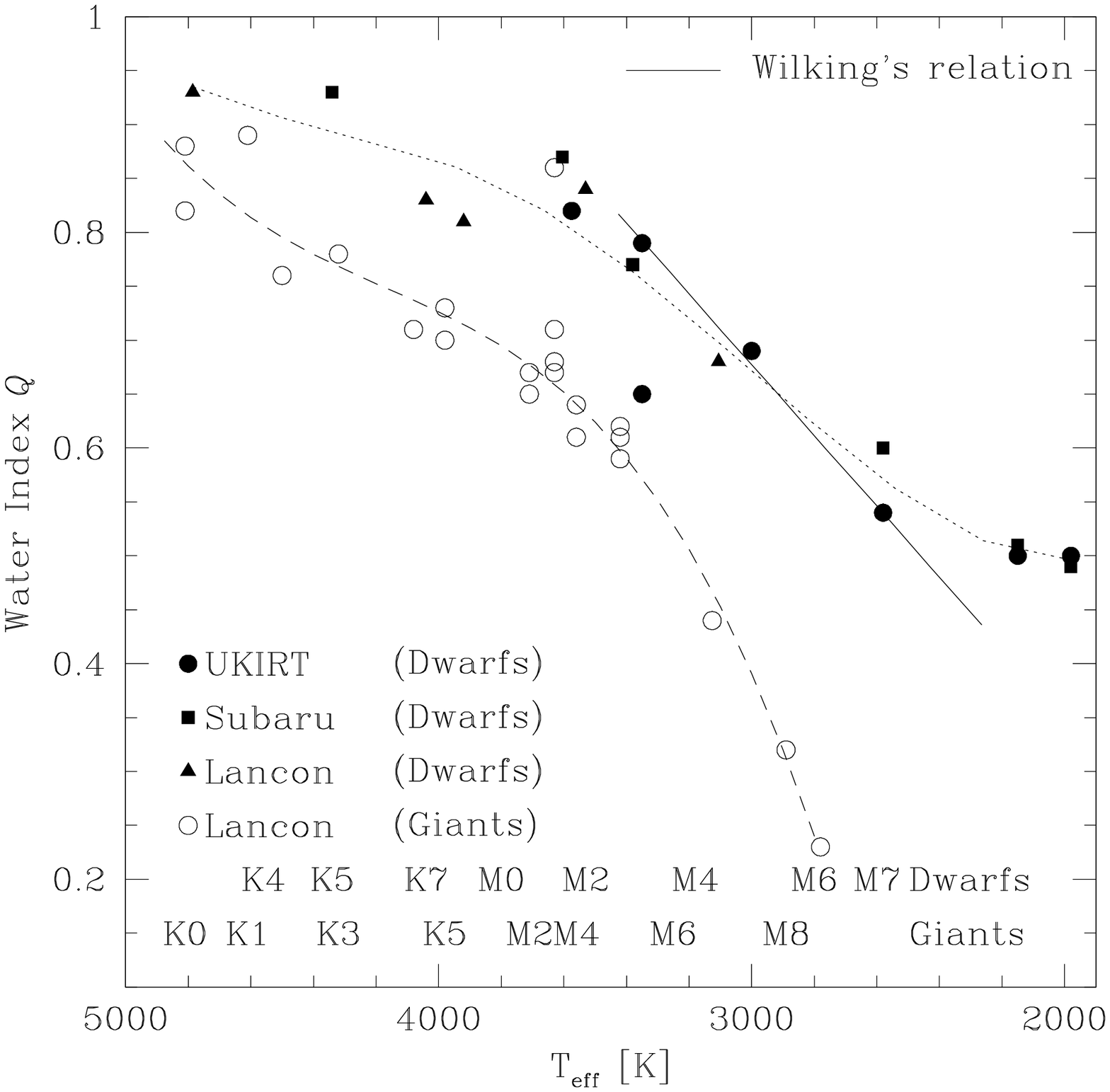}
  \end{center}
  \caption{The $Q$ index as a function of effective temperature. Data are
taken from our observations and from \citet{Lancon}. 
A solid line shows the relationship
between the $Q$ indices and spectral type for dwarfs \citep{Wilking}.
A dotted line shows a fitted line of the $Q$ index for the dwarfs, and
a dashed line for the giants.}
\label{Lancon_Q}
\end{figure}

The H$_{2}$O absorption bands in the near-infrared wavelengths are
a sensitive function of effective temperature for cool stars
(\cite{Aaronson}; \cite{Jones94}). 
Figure \ref{Lancon_Q} shows the
water index $Q$ of the dwarfs taken from our observations,
as well as those of dwarfs and giants taken from \citet{Lancon}.  
A dotted line represents a fitted line of the $Q$ index for the dwarfs
and a dashed line for the giants. In this figure, the relation between
the $Q$ index and spectral type for M dwarfs derived by \citet{Wilking}
also shown by a solid line.While Wilking's relation
is consistent with the plotted $Q$ index for the dwarfs,
the $Q$ indices for giants are much smaller than
the Wilking's relation.
Since the YSOs have surface gravity between dwarfs and giants, as we discuss
below, we should use both the relation for the dwarfs and that for the 
giants to estimate effective temperature of the YSOs.
Note that the veiling effect by a circumstellar disk could
increase the $Q$ index, leading to higher temperatures.

\subsubsection{Effective Temperatures of the YSOs}

The effective temperatures of the observed YSOs and their companions
are derived using 
the 2.26 $\micron$ / 2.21 $\micron$ feature ratio 
and the $Q$ index (T$_{\mathrm{eff}}$(NIR) in table \ref{teff_tbl}). 
The effective temperatures of most YSOs are less than 4000 K,
lower than that of typical CTTSs.

The effective temperatures of the observed CTTSs are also
derived from the near-infrared features. 
The effective temperatures estimated from the near-infrared spectroscopy are
consistent with those converted from optical spectral types for all the CTTSs.

\begin{table*}
\caption{Effective temperatures and spectral types of the YSOs and the CTTSs.}
\label{teff_tbl}
\begin{center}
{\small
\begin{tabular*}{183mm}{l@{\extracolsep{5pt}}
r@{\extracolsep{5pt}}
c@{\extracolsep{5pt}}
c@{\extracolsep{5pt}}
r@{\extracolsep{5pt}}
r@{\extracolsep{5pt}}
r@{\extracolsep{10pt}}
r@{\extracolsep{5pt}}
r@{\extracolsep{10pt}}
c}
\hline\hline
    Object &
T$_{\mathrm{eff}}(\mathrm{ratio})$\footnotemark[$*$] & 
T$_{\mathrm{eff}}(2.21)$\footnotemark[$\dagger$] & 
T$_{\mathrm{eff}}(2.26)$\footnotemark[$\ddagger$] &
T$_{\mathrm{eff}}(Q)$\footnotemark[$\S$] & 
T$_{\mathrm{eff}}$(NIR)\footnotemark[$\|$] & 
T$_{\mathrm{eff}}$(Opt)\footnotemark[$\#$] &
Sp.(V)\footnotemark[$**$] & 
Sp.(III)\footnotemark[$\dagger\dagger$] &
log Lbol\footnotemark[$\ddagger\ddagger$] \\
& [K] & [K] & [K] & [K] & [K] & [K] & & & [\LO] \\
\hline
ITG 2 & 2300$\sim$3400 & $<$5200 & $\cdots$& 2400$\sim$3200 
                & 2400$\sim$3200 & & M4$\sim$M8 & $>$M6 & $-1.11\sim-0.81$ \\
ITG 4 & field-star \\
ITG~5 & field-star \\
ITG 6 & 2300$\sim$3700& $\cdots$& $\cdots$& 3500$\sim$3700 
                & 3500$\sim$3700 & & M1$\sim$M2 & M2$\sim$M4 & $-1.72\sim-1.42$ \\
ITG 9A & $>$3000 & $\cdots$& $\cdots$& $>$3700 
                & $>$3700 & & $<$M1 & $<$M1 \\
ITG 9B & $\cdots$ & $\cdots$& $\cdots$& $>$3600 
                & $>$3600 & & $<$M2 & $<$M4 \\
ITG 13 & field-star \\
ITG 15A\&B& $>$3000 & $<$4800& $<$4700& 3000$\sim$3600 
                & 3000$\sim$3600 & & M2$\sim$M5 & M4$\sim$M7.5 & $-0.95\sim-0.65$ \\
ITG 15A& $>$3000 & $\cdots$& $\cdots$& 2900$\sim$3500
                & 3000$\sim$3500 & & M2$\sim$M5 & M5$\sim$M7.5 & $-0.95\sim-0.65$\\
ITG 15B& $>$3100 & $\cdots$& $\cdots$& $\cdots$
                & $\cdots$ & & $\cdots$ & $\cdots$ \\
ITG 17& $\cdots$ & $\cdots$& $\cdots$& 2700$\sim$3400 
                & 2700$\sim$3400 & & M3$\sim$M6.5 & $>$M5 & $-1.49\sim-1.19$ \\
ITG 18 & field-star \\
ITG 21& $>$2900 & $\cdots$& $\cdots$& 2600$\sim$3200 
                & 2900$\sim$3200 & & M4$\sim$M5 & M6$\sim$M7.5 & $-1.36\sim-1.06$ \\
ITG 24 & field-star \\
ITG 25A\&B& $>$3250 & $\cdots$& $\cdots$& 3600$\sim$4800 
                & 3600$\sim$4800 & & K3$\sim$M1 & K0$\sim$M2 & $-0.69\sim-0.39$ \\
ITG 25A& 2700$\sim$3600 & $\cdots$& $\cdots$& $>$3700
                & 2700$\sim$3600 & & M2$\sim$M6.5 & M4$\sim$M8 & $-0.57\sim-0.27$\\
ITG 25B& $\cdots$ & $\cdots$& $\cdots$& 2900$\sim$4400
                & 2900$\sim$4400 & & K5$\sim$M5 & K3$\sim$M8 & $-2.23\sim-1.93$\\
ITG 27& $>$2500 & $\cdots$& $\cdots$& 3600$\sim$4800 
                & 3600$\sim$4800 & & K3$\sim$M1 & K0$\sim$M2 & $-0.44\sim-0.14$\\
ITG 28 & field-star \\
ITG 29 & field-star \\
ITG 33A\&B& $>$3200 & $<$4400& $<$4400& 3300$\sim$4100 
                & 3300$\sim$4100 & & K7$\sim$M3 & K4$\sim$M6 & $-2.02\sim-1.72$ \\
ITG 33A& $\cdots$ & $\cdots$& $\cdots$& 3200$\sim$4000
                & 3200$\sim$4000 & & M0$\sim$M4 & M0$\sim$M6 & $-2.02\sim-1.72$\\
ITG 33B& $\cdots$ & $\cdots$& $\cdots$& $>$3700
                & $>$3700 & & $<$M1 & $<$M0 \\
ITG 36 & field-star \\
ITG 39 & field-star \\
ITG 40& 2000$\sim$3400& $<$4400& $\cdots$& 3000$\sim$3700    
                & 3000$\sim$3400 & & M3$\sim$M5 & M5$\sim$M7.5 & $-1.50\sim-1.20$ \\
ITG 41& $>$2750 & $\cdots$& $\cdots$& $>$3800   
                & $>$3800 & & $<$M0 & $<$M0 \\
ITG 43 & field-star \\
ITG 45A& field-star \\
ITG 45B& $\cdots$ & $\cdots$& $\cdots$& $>$3200
                & $>$3200 & & $<$M4 & $<$M6 \\
ITG 46& $>$2700 & $\cdots$& $\cdots$& 3500$\sim$4500 
                & 3500$\sim$4500 & & K4$\sim$M2 & K2$\sim$M4 & $-1.63\sim-1.33$\\

\hline
DD Tau& $>$2900 & $\cdots$& $\cdots$& 3500$\sim$4500
                & 3500$\sim$4500 & 3700$\sim$3900 & K4$\sim$M2 & K2$\sim$M4 \\
GH Tau& $>$2900 & $<$4800& $\cdots$& 3600$\sim$4800
                & 3600$\sim$4800 & 3500$\sim$3700 & K3$\sim$M2 & K0$\sim$M4 \\
FP Tau& $>$3100 & $<$4700& $<$5000& 3300$\sim$4100
                & 3300$\sim$4100 & 2900$\sim$3700 & K7$\sim$M3 & K4$\sim$M6 \\
\hline
\multicolumn{10}{@{}l@{}}{\hbox to 0pt{\parbox{180mm}{\footnotesize
\vspace*{2mm}
\footnotemark[$*$]
T$_{\mathrm{eff}}$ estimated from the 2.26 $\micron$ / 2.21 $\micron$ 
feature ratio.
    \par\noindent
\footnotemark[$\dagger$]
T$_{\mathrm{eff}}$ estimated from the the equi\-valent widths 
of the 2.21 $\micron$ feature. Because YSOs are subject to 
continuum emission and the strength of the line increases 
with lower temperature, only the upper limit of the effective 
temperature can be estimated.
    \par\noindent
\footnotemark[$\ddagger$]
T$_{\mathrm{eff}}$ estimated from the equi\-valent widths 
of the 2.26 $\micron$ feature.
\par\noindent
\footnotemark[$\S$]
T$_{\mathrm{eff}}$ estimated from the $Q$ 
index.
The range of temperatures for 
T$_{\mathrm{eff}}$(Q) includes estimates for both dwarf and 
giant surface gravities.
 These effective temperatures may be upper limits, 
due to veiling effect.
    \par\noindent
\footnotemark[$\|$]
T$_{\mathrm{eff}}$ estimated from 
the 2.26 $\micron$ / 2.21 $\micron$ feature ratio and the $Q$ index.
    \par\noindent
\footnotemark[$\#$]
T$_{\mathrm{eff}}$ converted from optical spectral type.
    \par\noindent
\footnotemark[$**$]
Estimated spectral type in the dwarf scale.
    \par\noindent
\footnotemark[$\dagger\dagger$]
Estimated spectral type in the giant scale.
    \par\noindent
\footnotemark[$\ddagger\ddagger$]
Bolometric luminosity for the midpoint of T$_{\mathrm{eff}}$(NIR),
estimated from the previous photometric study (see \S \ref{sec_bol}).
}\hss}}
\end{tabular*}
}
\end{center}
\end{table*}

\subsection{Luminosity Class}

\begin{figure}
  \begin{center}
    \FigureFile(85mm,85mm){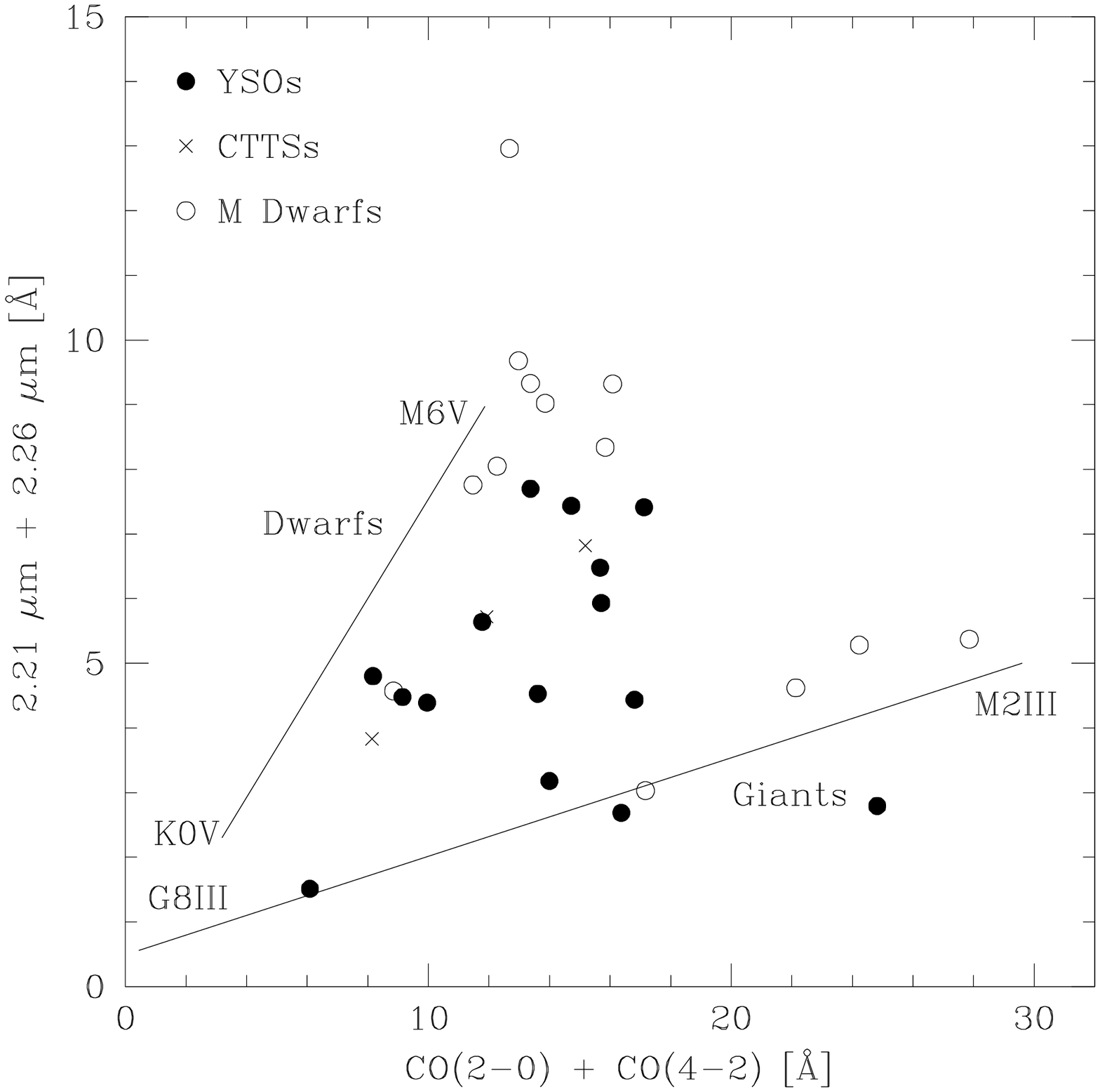}
  \end{center}
  \caption{Luminosity class of the YSOs. The combined equi\-valent width
of the 2.21 $\micron$ feature and the 2.26 $\micron$ feature is 
plotted against the combined CO band equi\-valent width.
Loci of dwarfs and giants \citep{Greene95} are shown by solid lines.}
\label{CONaCa}
\end{figure}

Since the 2.21 $\micron$ feature and the 2.26 $\micron$ feature 
are primarily sensitive to effective
temperature, and the depth of the CO band varies with surface gravity, 
the luminosity class of the object is estimated from a plot 
of the equi\-valent
widths of these atomic features against the widths of the CO band.
Figure \ref{CONaCa} shows the sum of the equi\-valent widths of the 2.21 $\micron$
feature and the 2.26 $\micron$ feature against the CO
band equi\-valent width of the YSOs, the CTTSs, and the late-type dwarfs. 
A sum of the equi\-valent widths of the CO (2-0) and CO (4-2) bands is
used as a
CO equi\-valent width. CO (3-1) band is not included because it
contains a sodium line \citep{Kleinmann}.
Relationships for giants and dwarfs \citep{Greene95}
are also shown in the figure by the solid lines.
Their fits are only applied to 
stars with spectral types $<$ M6V and $<$ M2III, and cannot be applied to
very late-type stars.
Points of very late-type dwarfs are close to the line for giants,
since their equi\-valent 
width of the 2.26 $\micron$ feature diminishes (see figure \ref{CaTeff}).

Most of the YSOs lie, like the CTTSs, between these two lines.
Therefore most of the YSOs and the CTTSs have a luminosity class 
between dwarfs and giants,
classified as luminosity class IV.

\subsection{Mass and Age of the YSOs}

Based on the effective temperature described above and the
bolometric luminosity described below,
the masses of the YSOs and their companions
are estimated with recent evolutionary tracks on the HR diagram.

\subsubsection{Bolometric Luminosity}
\label{sec_bol}
Bolometric
luminosity of the YSOs is estimated from the $J$-band
luminosity,
because the $J$-band emission arises primarily from the photosphere
of YSOs \citep{Bertout}. 
First, the $J$-band luminosity of the YSOs 
is corrected for interstellar extinction on the
color-color diagram (ITG).
Next, three cases are assumed, where the fraction of the photospheric
luminosity to the total luminosity is 1.0, 0.8, and 0.5 in the $J$-band.
Then, by comparing the photospheric luminosity of the YSOs 
with that of late-type dwarfs in the $J$-band \citep{Leggett},
the bolometric luminosity of the YSO's photosphere is estimated 

This procedure has the following uncertainties.
\begin{enumerate}
\item Extinction of each YSO is corrected on the color-color
diagram. If the extinction law of \citet{Bessell88} is used
instead of \citet{Koornneef}, a 6 \% larger value is derived as
visual extinction,
making the intrinsic luminosity brighter by about 2 \%.

\item Extinction is deduced from the distance between the observed 
color of the object and the
intrinsic color of CTTSs on the color-color diagram.
The intrinsic color of CTTSs deduced by \citet{Meyer97}
has a dispersion as much as 0.17 mag in $J-H$, corresponding
to an uncertainty of 1.5 mag in visual extinction or 0.16 dex in
bolometric luminosity.

\item The intrinsic color of YSOs might be different from the
intrinsic color of CTTSs. Central objects of the
YSOs are M type stars whereas those of CTTSs are mainly K
type stars. Color of an M type dwarf is about 0.1 mag redder than 
that of a K type dwarf both in $J-H$ and $H-K$.

\item We assumed that the distance to the Taurus molecular cloud is
140 pc. Hipparcos observations estimate 142 pc with 
an uncertainty of $\pm$ 14 pc as the distance to the cloud
\citep{Wichmann}. Therefore the uncertainty due to
distance is $\pm$ 0.1 dex in bolometric luminosity.

\item The fractions of the photospheric luminosity 
are assumed to be 1.0, 0.8, and 0.5 of the total luminosity in the $J$-band. 
In these three cases, uncertainty is 0.3 dex in bolometric luminosity.
If the fraction is smaller, photospheric luminosity is fainter.
Although the fraction value could be
deduced from veiling effect index $r_{J}$ \citep{Greene95},
$r_{J}$ has a large uncertainty for all YSOs due to
rather large observational uncertainties in the equi\-valent widths of the
metallic features. 
Therefore, we did not employ $r_{J}$ to estimate the photospheric luminosity.
\end{enumerate}

\subsubsection{HR diagrams}

\begin{figure}
  \begin{center}
    \FigureFile(85mm,85mm){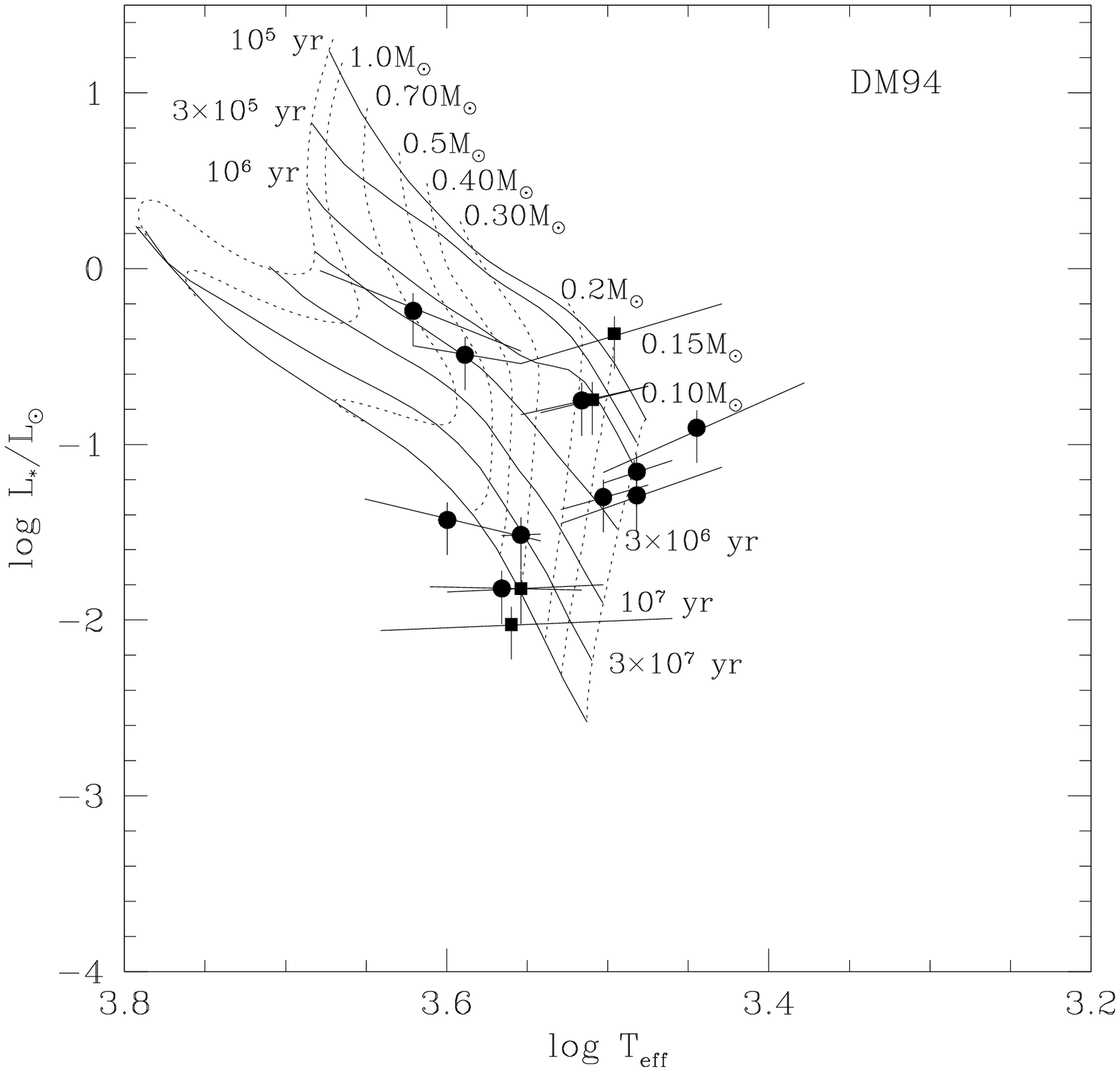}
  \end{center}
  \caption{The HR diagram of YSOs with
the evolutionary tracks of \citet{DM94} with the Alexander opacities and 
the CM convection overlaid.
Filled circles and filled squares show the loci of the YSOs deduced
from the UKIRT observations and from the Subaru observations, respectively.
T$_{\mathrm{eff}}$(NIR) in table \ref{teff_tbl} are adopted.
Error bars are tilted, because the bolometric luminosity is a function 
of the $J$-band luminosity and the effective temperature (see \S \ref{sec_bol}.}
\label{DM94}
\end{figure}

\begin{figure}
  \begin{center}
    \FigureFile(85mm,85mm){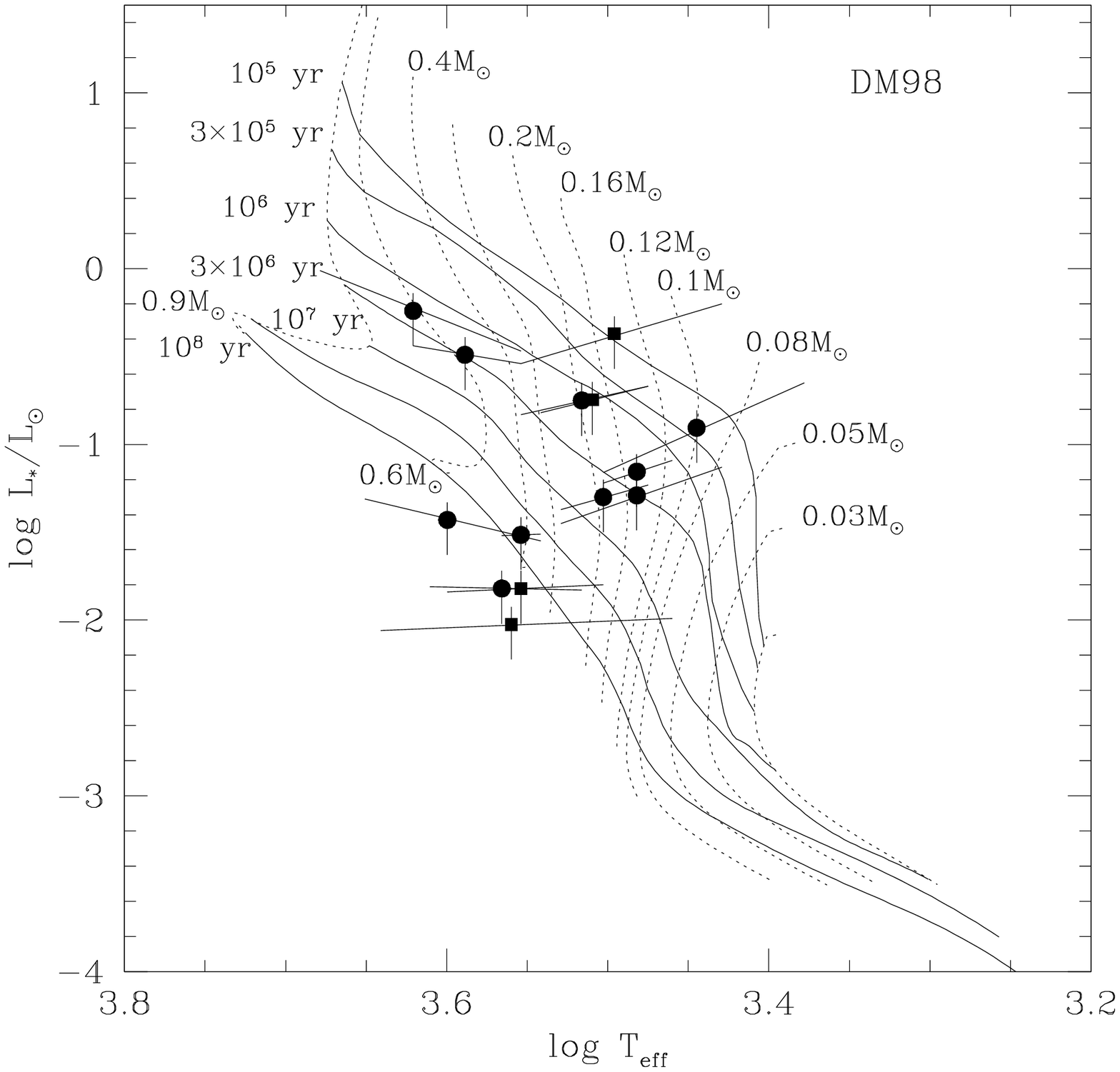}
  \end{center}
  \caption{The HR diagram of YSOs with evolutionary tracks
of \citet{DM98}
overlaid.}
\label{DM98}
\end{figure}

\begin{figure}
  \begin{center}
    \FigureFile(85mm,85mm){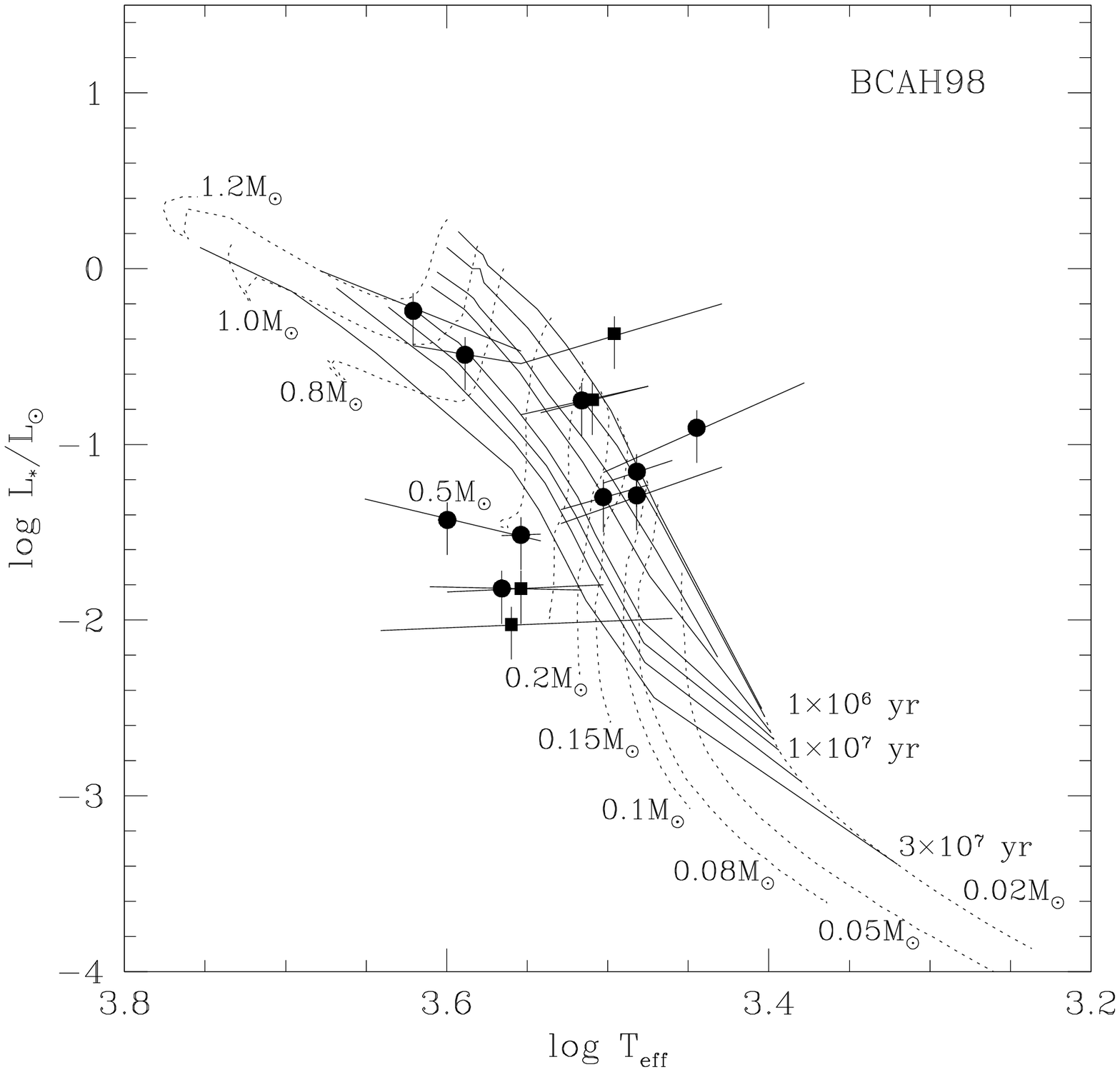}
  \end{center}
  \caption{The HR diagram of YSOs with evolutionary tracks
of \citet{Baraffe} overlaid.}
\label{B98}
\end{figure}

Several groups have recently developed and refined
evolutionary tracks for
low-mass stars, brown dwarfs, and even giant planets.
Figures \ref{DM94}, \ref{DM98}, and \ref{B98} show the HR diagrams
of the YSOs and their companions. Overlaid are the 
evolutionary tracks of 
\citet{DM94}, \citet{DM98},
and \citet{Baraffe}.

Mass and age for individual YSOs 
are estimated from
the HR diagrams with the evolutionary tracks (table \ref{mass_tbl}). 
Some YSOs are indeed very low-mass YSOs
(typically 0.1 \MO -- 0.3 \MO). 
{\it These objects are definitively a different population from CTTSs
in term of their mass.}
The uncertainties in the mass
are about a factor of 2 and those in the age are about 
a factor of 5 for the
YSOs as shown in the figures.
On the other hand, none of the mass and age of the companions are determined
mainly due to the poor signal-to-noise ratio on the spectra.
Additional notes are described in the appendix for most companions.

\begin{figure}
  \begin{center}
    \FigureFile(85mm,85mm){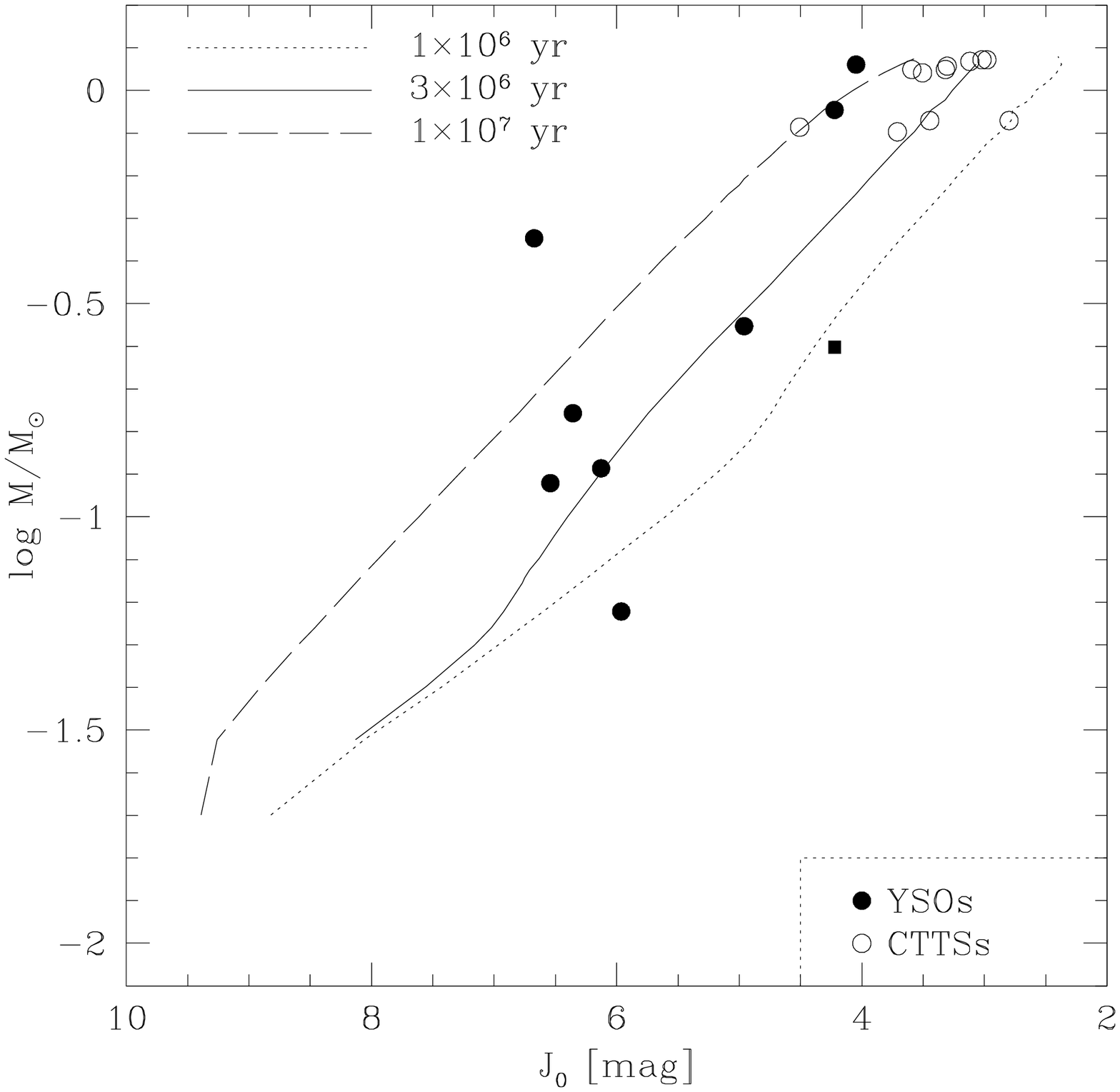}
  \end{center}
  \caption{Mass -- absolute $J$-band magnitude relation of 
the YSOs and the CTTSs listed in \citet{Strom88}.
The $J$-band magnitudes were corrected for interstellar extinction on 
the color-color diagram (ITG).
The mass of the YSOs and the CTTSs are estimated from the
evolutionary track of \citet{Baraffe}.
The distance modulus of the Taurus molecular cloud is
5.73 \citep{Elias}.}
\label{jmass}
\end{figure}

Mass-luminosity relation of the YSOs
is shown in figure \ref{jmass}, where also shown
are the mass-luminosity relations of CTTSs \citep{Strom88}
and the predicted relations
for three ages calculated from  
\citet{Baraffe}.
Most of the YSOs are
distributed within one order of magnitude in the ages.
Both CTTSs and the YSOs have
near-infrared excesses, implying existence of circumstellar
disks. 
The ages of these objects indicate the survival time of
the circumstellar disk, and it is interesting to ask whether the mass of
the central object affects the evolution of circumstellar disk.
The ages of most YSOs are less than 10$^{7}$ yr, similar
to the age of CTTSs. 
However, 
since the mass--age relation of YSOs depends strongly on evolutionary
tracks (figures \ref{DM94},
\ref{DM98}, \ref{B98}, and table \ref{mass_tbl}), we cannot conclude
that disk survival time is independent on the mass of the central
object.

The small amount of objects prevents us from making a mass function
of the YSOs. In the four other regions but except for Heiles Cloud 2,
\citet{Luhman00} shows that the mass function
has a peak at 0.8 \MO and is relatively flat between 0.1 \MO and 
0.8 \MO. \citet{Briceno98}, as well as \citet{Luhman00}, suggest a deficit of
brown dwarfs in the regions, although \citet{Martin01} have identified four
brown dwarfs in two sparsely populated stellar groups. 
Given a limiting magnitude
of the ITG survey ($K=13.4$) and an age of 1 Myr, we expect to reach a
mass limit of 0.03 \MO for $Av=0$ with the survey. In order to take a census of
brown dwarfs in Heiles cloud 2 and to construct an initial mass function
of the cloud, a deeper spectroscopic search is required.

\begin{table*}
  \caption{Mass and age of the YSOs.}\label{mass_tbl}
  \begin{center}
    \begin{tabular}{lrrrrrr}
\hline\hline
    Object &
\multicolumn{2}{c}{DM94\footnotemark[*]}      &
\multicolumn{2}{c}{DM98\footnotemark[$\dagger$]}      &
\multicolumn{2}{c}{BCAH98\footnotemark[$\ddagger$]}      \\
& mass & age & mass & age & mass & age  \\
& [\MO] & [10$^{6}$ yr] & [\MO] & [10$^{6}$ yr] & [\MO] & [10$^{6}$ yr] \\
\hline
ITG 2   & $<$0.15 
        & $<$2
        & 0.05 -- 0.18
        & $<$3
        & $<$0.2
        & $<$3  \\
ITG 6   & 0.25 -- 0.4 
        & 20 -- 50\footnotemark[\S]
        & 0.35 -- 0.45
        & $>$30\footnotemark[\S]
        & 0.4 -- 0.5
        & $>$50\footnotemark[\S]    \\
ITG 15A\&B  & 0.1 -- 0.4 
        & 0.1 -- 3
        & 0.12 -- 0.4
        & 0.2 -- 5
        & $<$0.55
        & $<$8    \\
ITG 15A & 0.1 -- 0.3 
        & 0.1 -- 3
        & 0.12 -- 0.3
        & 0.2 -- 4
        & $<$0.45
        & $<$5    \\
ITG 17  & $<$0.2 
        & $<$10
        & 0.08 -- 0.3
        & 0.3 -- 20
        & $<$0.3
        & $<$20    \\
ITG 21  & $<$0.15
        & $<$2
        & 0.11 -- 0.19
        & 1 -- 5
        & $<$0.2
        & $<$4    \\
ITG 25A\&B  & 0.35 -- 0.8
        & 1 -- 10
        & 0.35 -- 0.75
        & 2 -- 8
        & 0.57 -- 1.00
        & 4 -- 18    \\
ITG 25A & $<$0.35
        & $<$1
        & 0.09 -- 0.35
        & $<$2
        & $<$0.57
        & $<$4    \\
ITG 25B & $\cdots$
        & $\cdots$
        & $>$0.08
        & $>$10
        & $>$0.06
        & $>$4.5    \\
ITG 27  & 0.3 -- 1.2
        & 1 -- 5
        & $>$0.3
        & 1 -- 8
        & 0.57 -- 1.2
        & 3 -- 18          \\
ITG 33A\&B  & $>$0.15
        & $>$10\footnotemark[\S]
        & $>$0.2
        & $>$30\footnotemark[\S]
        & $>$0.2
        & $>$30\footnotemark[\S]    \\
ITG 33A & $>$0.15
        & $>$10\footnotemark[\S]
        & $>$0.2
        & $>$30\footnotemark[\S]
        & $>$0.14
        & $>$18\footnotemark[\S]    \\
ITG 40  & $<$0.2 
        & 1 -- 7
        & 0.14 -- 0.3
        & 2 -- 15
        & 0.1 -- 0.32
        & 4 -- 18    \\
ITG 41  & $\cdots$ 
        & $\cdots$ 
        & $\cdots$ 
        & $\cdots$ 
        & $\cdots$ 
        & $\cdots$    \\
ITG 46  & $>$0.25 
        & $>$20\footnotemark[\S]
        & $>$0.35
        & $>$30\footnotemark[\S]
        & $>$0.35
        & $>$30\footnotemark[\S] \\
\hline
\multicolumn{7}{@{}l@{}}{\hbox to 0pt{\parbox{180mm}{\footnotesize
\vspace*{2mm}
    \footnotemark[$*$]
\citet{DM94} with the Alexander opacities and the CM convection
    \par\noindent
    \footnotemark[$\dagger$] \citet{DM98}
    \par\noindent
    \footnotemark[$\ddagger$] \citet{Baraffe}
    \par\noindent
    \footnotemark[\S] These ages are discussed further in the appendix.
    }\hss}}
    \end{tabular}
  \end{center}
\end{table*}

\subsection{Comparison with other studies}

\citet{Wilking} carried out $K$-band spectroscopic observations of
low-luminosity sources in the $\rho$ Oph cloud. They
estimated spectral types of the objects using the $Q$ index,
then derived effective temperatures using the conversion for dwarfs.
The effective
temperature derived from their method is the same value as the lower
limit of effective temperatures derived from the $Q$ index in this
paper. This is because M giants have a higher temperature than M
dwarfs even with the same $Q$ index, due to gravity dependence of the
water band depth (see figure \ref{Lancon_Q}). 
For example, ITG 40 has a $Q$ of 0.67, which corresponds to a spectral type of
M4.7 for Wilking's relation. This spectral type corresponds to 2980
K in the dwarf scale, which is consistent with the lower limit on
the effective temperature derived from the $Q$ index in this
paper.

They estimated extinction of each object from the $J - H$ color,
assuming no infrared excesses in the $J$- and $H$-bands.
This procedure could lead to a slightly higher value for extinction 
than our method. For example
$Av$ of ITG 40 is estimated to be 20.0 mag by their method. On
the other hand, our value is 16.7 mag. Larger $Av$ leads to higher
values in bolometric luminosity. 
Bolometric luminosity of ITG 40
is estimated to be 0.1 \LO ~by their method, whereas 0.05 \LO ~by us,
even though discrepancy is within uncertainty.

\citet{LuhmanRieke} carried out $K$-band spectroscopy of
low-luminosity sources in the L 1495 cloud in Taurus. 
They derived the spectral type of the object by fitting the 2.21 $\micron$
feature and 2.26 $\micron$ strengths of the object to those of dwarf stars
varying in amounts of veiling.
With their method, spectral type of ITG 40 is estimated to be M3
or M7. Derived spectral type corresponds to 2600 K $\sim$
3400 K, in agreement with our result.


\section{Conclusions}
\begin{enumerate}

\item We have carried out
near-infrared spectroscopic observations of 23 very low-luminosity YSO 
candidates and 5 of their companions in Heiles Cloud 2 in the
Taurus molecular cloud.
Near-infrared spectroscopy is essential to characterize
objects in the color-color diagram.
Out of the 28 objects, 5 objects have Br$\gamma$ in emission
and 7 have "flat" spectra over the Br$\gamma$ feature. 
We conclude that these 12 objects are indeed YSOs.

\item Compiling near-infrared spectra of dwarfs and giants taken
by us as well as those in the literature, the ratio of the 2.26 $\micron$
feature to the 2.21 $\micron$ feature
turns out to be a good indicator of the effective
temperature for M type stars.

%

\item The effective temperatures of the YSOs are 
determined. These objects are cool (T$_{\mathrm{eff}} <$ 4000 K)
YSOs.

\item The mass of these YSOs is 
estimated from the HR diagram with recent
evolutionary tracks. Some objects are
very low-mass (0.1 \MO -- 0.3 \MO) YSOs.

\item The age of these YSOs appears to be 10$^{5}$ --
10$^{7}$ years. However, the deduced age depends on
evolutionary track models.

%

\end{enumerate}

~\\

~\\
~\\

We are grateful to T. Geballe, T. Kerr,
and Y. Oasa for help with the UKIRT observations, and H. Terada, N. Kobayashi, 
and B. Potter for the Subaru observations.
We thank T. Tsuji, and T. Nakajima for
discussions on spectral features of late type stars. 
We also thank our referee, B. Wilking, for many helpful comments.
Y. I. is supported from the Sumitomo Foundation.
A part of this study was supported by the UK-Japan collaboration
fund from the JSPS.
The United Kingdom Infrared Telescope is operated by the Joint
Astronomy Centre on behalf of the U.K. Particle Physics and
Astronomy Research Council.
Subaru Telescope is operated by the National Astronomical Observatory of Japan.

\appendix
\section*{Individual Objects}

\subsection*{ITG 2}

Mass of this object is estimated to be less than 0.2 \MO ~by
any evolutionary track, possibly a young brown dwarf.
Age of this object is estimated to be less than $3\times10^{6}$.

\subsection*{ITG 6 (GM Tau)}

\citet{Gomez} derived the magnitude of this star as
8.06 in the $K$-band, one order of magnitude brighter than the magnitude derived
by ITG and by \citet{Kenyon}. If the bolometric luminosity is one
order of magnitude brighter, the age of this star is estimated to be 
around $10^{6}$ yr,
while the mass of the star is still about 0.3 \MO.

\citet{Kenyon} classified spectral type of this star as "continuum".
Therefore the spectrum of this object has a large amount of veiling.
Since only the lower limit of the effective temperature is estimated 
from the $Q$ index, 
true effective temperature might be lower than that estimated here.
If half of the $K$-band flux comes from a circumstellar disk, 
for example, 
the $Q$ index changes from 0.81
to 0.62 and the effective temperature to 2700 K $\sim$ 3500 K.
This value is not inconsistent with the effective temperature deduced from
the ratio of the 2.26 $\micron$ feature to the 2.21 $\micron$ feature.
With this temperature, the mass and the age of the object would be 0.1 
\MO ~and $3\times10^{6}$ yr.

\subsection*{ITG 9}

The fact that both spectra of the binary have neither the strong metallic 
features nor the deep water absorption bands indicates high effective
temperatures of the objects. They both might be field-stars.

\subsection*{ITG 15A}

The effective temperatures deduced from the UKIRT observations and that from
the Subaru observations are very consistent with each other.

\subsection*{ITG 25 (IRAS 04370+2559)}

While 
the UKIRT observations and the Subaru observations are consistent with
each other in the strength of the 2.26 $\micron$ feature as well as the
$Q$ index, there is a discrepancy
in the strength of the 2.21 $\micron$ feature, resulting in a 
different effective
temperature. Because of higher signal-to-noise ratio in the Subaru observations
than in the UKIRT observations, we use the effective temperature
 derived from the Subaru
observations. For the Subaru observations, effective temperatures derived
from the feature ratio and from the $Q$ index are also different.
As is the case of ITG 6, 
the veiling effect can easily increase the $Q$ index, leading to 
a higher effective temperature.
Because this object is an IRAS source indicating
its young age, we believe the effective temperature derived from
the feature ratio rather than that from the $Q$ index.

For consistency between the effective temperature derived from the feature
ratio and that from the $Q$ index, the $Q$ index should be 0.65, which 
means that 70\% of the $K$-band luminosity is not of photospheric origin but 
from a circumstellar disk.
If the same portion of the luminosity is due to the veiling for the
companion ITG 25B, the $Q$ index of the companion is suppressed to 0.08.
This very low value indicates that effective temperature is less than
2700 K and the age is less than $1\times10^{6}$ yr, 
implying a low-mass brown dwarf.

\subsection*{ITG 33}

\citet{Martin00} classified the primary as a T Tauri star with spectral type
of M3 and with $A_{v}$ of 3.5 mag from optical spectroscopy.
Spectral type derived from near-infrared spectroscopy is M0 or M4, 
consistent with the optical spectroscopy. Visual extinction is estimated
to be 3.94$\pm$0.55 from near-infrared photometry, also consistent with the
optical spectroscopy. \citet{Martin00} proposed an edge-on disk T Tauri star,
like a companion of HK Tau (\cite{Koresko98}; \cite{Stapelfeldt98}). 
If so, visual extinction would
be very large as much as 50 mag, like the HK Tau companion,
inconsistent with $A_{v}$ for ITG 33A.
The location of ITG 33A is odd as a YSO in the HR diagram.
Veiling effect may lead to high effective temperature. A spectrum with
high signal-to-noise ratio is required to estimate effective temperature
from metallic features.

\citet{Martin00} also classified ITG 33B as a field-star from
optical spectroscopy.
However, their optical spectrum, as well as our near-infrared spectrum, are
too noisy to classify this object as a field-star or a featureless YSO.
Higher signal-to-noise ratio spectrum at optical and/or of near-infrared
wavelength is required.

\subsection*{ITG 41 (IRAS 04385+2550)}

This object is identified at optical wavelengths as Haro 6-34.
Weak absorption features which may imply large amount of veiling
prevents us from estimating the effective temperature, thus its mass and age.

\subsection*{ITG 45}

\citet{Martin00} found an H$\alpha$ absorption line in a spectrum of 
ITG 45A, indicating a field-star. Near-infrared spectrum of the primary
contains Br$\gamma$ absorption, also indicating a field-star.
\citet{Martin00} claims that the companion, ITG 45B, is also a field-star.
However, the optical spectrum, as well as our near-infrared spectrum, is
too noisy to classify it as a field-star or a YSO.

\subsection*{ITG 46}

The location of ITG 46 in the HR diagram is unusual. Again a large mount of 
veiling may lead to high effective temperature. Otherwise, this object
may be a field-star.


\end{document}